\documentstyle[12pt]{article}
\begin{document}

\hspace*{63ex} OCU-PHYS-180 

\hspace*{63ex} January 2001 

\begin{center}
{\Large {\bf 
Gauge-boson propagator in out of equilibrium quantum-field system 
and the Boltzmann equation
}} \\ 

\hspace*{3ex}

\hspace*{3ex}

\hspace*{3ex}

{\large {\sc A. Ni\'{e}gawa}\footnote{E-mail: 
niegawa@sci.osaka-cu.ac.jp}

{\normalsize\em Department of Physics, Osaka City University } \\ 
{\normalsize\em Sumiyoshi-ku, Osaka 558-8585, JAPAN} } \\ 

\hspace*{3ex}

\hspace*{3ex} 

\hspace*{3ex}

{\large {\bf Abstract}} \\ 
\end{center} 
We construct from first principles a perturbative framework 
for studying nonequilibrium quantum-field systems that include 
gauge bosons. The system of our concern is quasiuniform 
system near equilibrium or nonequilibrium quasistationary system. We 
employ the closed-time-path formalism and use the so-called gradient 
approximation. No further approximation is introduced. We construct 
a gauge-boson propagator, with which a well-defined perturbative 
framework is formulated. In the course of construction of the 
framework, we obtain the generalized Boltzmann equation (GBE) that 
describes the evolution of the number-density functions of 
gauge-bosonic quasiparticles. The framework allows us to compute 
the reaction rate for any process taking place in the system. 
Various processes, in turn, cause an evolution of the systems, which 
is described by the GBE. 
\newpage
%%%%%%%%%%%%%%%%%%%%%%%%%%%%%%%%%%%%%%%%%%%%%%
%%%% SECTION I %%%%%%%%%%%%%%%%%%%%%%%%%%%%%%%
%%%%%%%%%%%%%%%%%%%%%%%%%%%%%%%%%%%%%%%%%%%%%%
\section{Introduction} 
Ultrarelativistic heavy-ion-collision experiments at the BNL 
Relativistic Heavy Ion Collider (RHIC) have begun and will soon 
start at the CERN Large Hadron Collider (LHC) in an anticipation of 
producing a quark-gluon plasma (QGP) [see, e.g., \cite{bjo,le-b}]. 
The QGP to be produced is an expanding nonequilibrium system. 
Studies of the QGP as such have just begun. 

In previous papers, a perturbative framework has been formulated 
{\em from first principles} for dealing with out-of-equilibrium 
complex-scalar field system \cite{nie}, $O(N)$ linear-sigma 
system \cite{nie1}, and the system that includes massless fermions 
\cite{nie2}. In this paper, we take up the out-of-equilibrium 
quantum-field theories that includes gauge bosons. We deal with the 
gauge-boson and the FP-ghost sectors in Coulomb gauge and, following 
the same procedure as in \cite{nie,nie1,nie2}, we construct the 
gauge-boson and FP-ghost propagators and, thereby, frame a 
perturbation theory. Only approximation we employ is the so-called 
gradient approximation (see below). We use the closed-time-path 
(CTP) formalism \cite{sch,chou,lan} of nonequilibrium statistical 
quantum-field theory. It turns out that the form of gauge-boson 
propagator is quite complicated. 

Throughout this paper, we are interested in quasiuniform systems 
near equilibrium or nonequilibrium quasistationary systems. Such 
systems are characterized by two different spacetime scales 
\cite{chou}; microscopic or quantum-field-theoretical and 
macroscopic or statistical. The first scale, the microscopic 
correlation scale, characterizes the reactions taking place in the 
system, while the second scale measures the relaxation of the 
system. For a weak coupling theory, in which we are interested in 
this paper, the former scale is much smaller than the latter 
scale.\footnote{It should be noted, however, that, as the system 
approaches the critical point of the phase transition, the 
microscopic correlation scale diverges. Thus, the formalism 
developed in this paper applies to the systems away from the 
critical point.} In a derivative expansion of some quantity $F$ with 
respect to macroscopic spacetime coordinates $X$, we use the 
gradient approximation throughout: 
%%%%%%%%%%%%%%%%%%%%%%%%%%%%%%%%%%%%%%%%%%%%%%%%%%%%%%%%%%
\begin{equation} 
F (X, ...) \simeq F (Y, ...) + (X - Y)^\mu \partial_{Y^\mu} F (Y, 
...) \, . 
\label{gra} 
\end{equation} 
%%%%%%%%%%%%%%%%%%%%%%%%%%%%%%%%%
Let $\Delta (x, y)$ be a generic propagator. For the system of our 
concern, $\Delta (x, y)$, with $x - y$ fixed, does not change 
appreciably in $(x + y) / 2$. We refer to the first term on the 
right-hand side (RHS) of Eq.~(\ref{gra}) as the {\em leading part 
(term)} and to the second term as the {\em gradient part (term)}. 
The self-energy part $\Pi (x, y)$ enjoys a similar property. 
Thus, as usual, we choose the relative coordinates $x - y$ as the 
microscopic 
coordinates while $X \equiv (x + y) / 2$ as the macroscopic 
coordinates. A Fourier transform with respect to $x - y$ yields 
%%%%%%%%%%%%%%%%%%%%%%%%%%%%%%%%%%%%%%%%%%%%%%%%%%%%%%%%%%
\begin{equation} 
\Delta (x, y) = \int \frac{d^{\, 4} P}{(2 \pi)^4} \, 
e^{- i P \cdot (x - y)} \Delta (X; P) \mbox{\hspace*{6ex}} 
(X \equiv (x + y) / 2) 
\label{Fou} 
\end{equation} 
%%%%%%%%%%%%%%%%%%%%%%%%%%%%%%%%%
$(P^\mu = (p^0, {\bf p}))$ together with a similar formula for 
$\Pi$. The above observation shows that $P^\mu$ in 
Eq.~(\ref{Fou}) can be regarded as the momentum of the quasiparticle 
participating in the microscopic reaction under consideration. We 
shall freely use $\Delta (x, y)$ or $\Delta (X; P)$ [$\Pi (x, y)$ 
or $\Pi (X; P)$], which we simply write $\Delta$ [$\Pi$] whenever 
obvious from the context. 

The perturbative framework to be constructed accompanies the 
generalized Boltzmann equation (GBE) for the number density of 
quasiparticles. The framework allows us to compute any 
reaction rate by using the reaction rate formula \cite{noo}. 
Substituting the computed net production rates of 
quasiparticles into the GBE, one can determine the number 
densities as functions of macroscopic spacetime coordinates, which 
describes the evolution of the system. 

For definiteness, we take up QCD. However, the procedure has little 
to do with QCD and then the framework to be constructed may be used 
for any theory that includes gauge boson(s) with almost no 
modification. The plan of the paper is as follows: In Sec.~II, on 
the basis of CTP formalism, we construct the bare gauge-boson and 
FP-ghost propagators, with which the \lq\lq bare-$N$ scheme'' may be 
constructed. In Sec.~III, we set up the basis for formulating the 
\lq\lq physical-$N$ scheme.'' In Sec.~IV, we make up the 
self-energy-part resummed propagators. In Sec.~V, we impose the 
condition that no large contribution appears in perturbative 
calculation, so that a \lq\lq healthy'' perturbative framework is 
constructed. It is shown that, on the energy-shell, the condition 
turns out to the generalized Boltzmann equation. In Sec.~VI, we 
frame a concrete perturbative framework. Section VII is devoted to 
summary and outlook. Concrete derivation of various formula used in 
the text is made in Appendices. 
%%%%%%%%%%%%%%%%%%%%%%%%%%%%%%%%%%%%%%%%%%%%%%
%%%% SECTION I %%%%%%%%%%%%%%%%%%%%%%%%%%%%%%%
%%%%%%%%%%%%%%%%%%%%%%%%%%%%%%%%%%%%%%%%%%%%%%
\section{Closed-time-path formalism} 
\subsection{Preliminary} 
For definiteness, we take up QCD and adopt the Coulomb gauge. We 
deal only with the gluon and the FP-ghost sectors in an out of 
equilibrium QCD system. Singling out the free parts of the gluon and 
of the FP-ghosts, we write the Lagrangian (density) as 
%%%%%%%%%%%%%%%%%%%%%%%%%%%%%%%%%%%%%%%%%%%%%%%%%%%%%%%%%%%%%%%%
\begin{eqnarray} 
&& {\cal L} (A^{(a) \mu} (x), \eta^{(a)} (x), \bar{\eta}^{(a)} (x), 
...) = - \frac{1}{2} A^{(a) \mu} {\cal D}_{\mu 
\nu} A^{(a) \nu} + \bar{\eta}^{(a)} \nabla^2_{\bf X} \eta^{(a)} + 
... \, , \nonumber \\ 
&& {\cal D}^{\mu \nu} \equiv - g^{\mu \nu} \partial^2 + 
\partial^\mu \partial^\nu - \frac{1}{\alpha} \left[ (\partial^\mu 
- \partial^0 n^\mu) (\partial^\nu - \partial^0 n^\nu)  \right] \, . 
\label{Ddesu} 
\end{eqnarray} 
%%%%%%%%%%%%%%%%%%%%%%%%%%%%%%%%%%%%%%%%%%%%%%%%%%%%%%%%%%%%%%%%
Here $n^\mu = (1, {\bf 0})$, $A^{(a) \mu}$ (\lq\lq $a$'' the color 
index) is the gluon fields, and $\eta^{(a)}$ and $\bar{\eta}^{(a)}$ 
are the FP-ghost fields. The CTP formalism is formulated 
\cite{sch,chou,lan} by introducing an oriented closed-time path $C$ 
$(= C_1 + C_2)$ in a complex-time plane, that goes from $- \infty$ 
to $+ \infty$ $(C_1)$ and then returns from $+ \infty$ to $- \infty$ 
$(C_2)$. The real time formalism is achieved by doubling every 
degree of freedom: For example, for gauge fields, $A^{(a) \mu} \to 
(A^{(a) \mu}_1, A^{(a) \mu}_2)$, where $A^{(a) \mu}_1 (x_0, {\bf x}) 
= A^{(a) \mu} (x_0, {\bf x})$ with $x_0 \in C_1$ and $A^{(a) \mu}_2 
(x_0, {\bf x}) = A^{(a) \mu} (x_0, {\bf x})$ with $x_0 \in C_2$. A 
field with suffix \lq $i$' $(i = 1, 2)$ is called a type-$i$ field. 
One can introduce a two-dimensional space, in which every field is 
a \lq\lq vector'' whose first (second) component is the type-1 
(type-2) field. In equilibrium thermal field theory, this space is 
called the thermal space, so that we use this terminology in the 
following. The classical contour action that governs the dynamics of 
nonequilibrium systems is written in the form 
%%%%%%%%%%%%%%%%%%%%%%%%%%%%%%%%%%%%%
\begin{eqnarray} 
&& \int_C d x_0 \int d {\bf x} \, {\cal L} (A^{(a) \mu} (x), 
\eta^{(a)} (x), \bar{\eta}^{(a)} (x), ...) = \int_{- \infty}^{+ 
\infty} d x_0 \int d 
{\bf x} \, \hat{\cal L} (x) \, , \nonumber \\ 
&& \hat{\cal L} (x) \equiv {\cal L} (A^{(a) 
\mu}_1 (x), \eta^{(a)}_1 (x), \bar{\eta}^{(a)}_1 (x), ...) - 
{\cal L} (A^{(a) \mu}_2 (x), 
\bar{\eta}^{(a)}_2 (x), \bar{\eta}^{(a)}_2 (x), ...) \, , 
\label{hat-L} 
\end{eqnarray} 
%%%%%%%%%%%%%%%%%%%%%%%%%%%%%%%%%%%%%%%%%%%%%
where \lq\lq $...$'' stands for quark fields. $\hat{\cal L}$ here is 
sometimes called a hat-Lagrangian [cf. \cite{ume}]. Throughout this 
paper, we do not deal with initial correlations (see, e.g., 
\cite{chou}). 
%%%%%%%%%%%%%%%%%%%%%%%%%%%%%%%%%%%%%%%%%%
\subsubsection{Gluon sector} 
Following standard procedure \cite{chou}, the four kind 
of gluon propagators emerges: 
%%%%%%%%%%%%%%%%%%%%%%%%%
\begin{eqnarray} 
\left( \Delta_{1 1}^{(a b) \mu \nu} (x, y) \right) & \equiv & - i 
\mbox{Tr} \left[ T \left\{ A_1^{(a) \mu} (x) A_1^{(b) \nu} 
(y) \right\} \rho \, \right] \, , \nonumber \\ 
\left( \Delta_{1 2}^{(a b) \mu \nu } (x, y) \right) & \equiv & - i 
\mbox{Tr} \left[ A_2^{(b) \nu} (y) A_1^{(a) \mu} (x) \, 
\rho \, \right] \, , \nonumber \\ 
\left( \Delta_{2 1}^{(a b) \mu \nu} (x, y) \right) & \equiv & - i 
\mbox{Tr} \left[ A^{(a) \mu}_2 (x) A_1^{(b) \nu} (y) \, \rho \, 
\right] \, , \nonumber \\ 
\left( \Delta_{2 2}^{(a b) \mu \nu} (x, y) \right) & \equiv & - i 
\mbox{Tr} \left[ \overline{T} \left\{ A_2^{(a) \mu} (x) A_2^{(b) 
\nu} (y) \right\} \rho \, \right] \, , 
\label{kore} 
\end{eqnarray} 
%%%%%%%%%%%%%%%%%%%%%%%%%%%%%%%%%%%%%%%%%%%%%%
where $\rho$ is the density matrix, and $T$ ($\overline{T}$) is the 
time-ordering (antitime-ordering) symbol. We restrict 
ourselves to the case where 
$\rho$ is color singlet, so that $\Delta_{i j}^{(a b) \mu \nu} = 
\delta^{a b} \Delta_{i j}^{\mu \nu}$. Throughout in the sequel, we 
drop the color index. At the end of calculation we set $A_1^\mu = 
A_2^\mu$ \cite{chou}. Let us use bold-face letter ${\bf \Delta}_{i 
j} (x, y)$ for denoting $4 \times 4$ matrix, whose $(\mu \, 
\nu)$-component is $\Delta_{i j}^{\mu \nu} (x, y)$. We further 
introduce a matrix propagator $\hat{\bf \Delta} (x, y)$, where the 
\lq caret' denotes the $2 \times 2$ matrix in thermal space, whose 
$(i \, j)$-component is ${\bf \Delta}_{i j} (x, y)$. The matrix 
self-energy part $\hat{\bf \Pi} (x, y)$ is defined similarly. 

Deduction of the gluon propagator $\hat{\bf \Delta}$ is carried 
out below in subsection B. 
%%%%%%%%%%%%%%%%%%%%%%%%%%%%%%%%%%%%%%%%%%%%%%%%%%%%%%%%%%
\subsubsection{FP-ghost sector} 
The FP-ghost fields are unphysical fields, so that they are absent 
in the system under consideration. Then, the FP-ghost propagators 
are the same as in vacuum theory. This means that the $2 \times 2$ 
matrix propagator that act on a thermal space is diagonal, whose 
diagonal elements are 
%%%%%%%%%%%%%%%%%%%%%%%%%
\begin{eqnarray*} 
\left[ \Delta_g^{(a b)} (x, y) \right]_{1 1} & \equiv & - i 
\langle 0 | T \left\{ \eta_1^{(a)} (x) \bar{\eta}_1^{(b)} (y) 
\right\} | 0 \rangle \, , \nonumber \\ 
\left[ \Delta_g^{(a b)} (x, y) \right]_{2 2} & \equiv & - i \langle 
0 | \overline{T} \left\{ \eta_2^{(a)} (x) \bar{\eta}_2^{(b)} (y) 
\right\} | 0 \rangle \, . 
\end{eqnarray*} 
%%%%%%%%%%%%%%%%%%%%%%%%%%%%%%%%%%%%%%%%%%%%%%
In what follows, as in the gluon sector, we drop the color indices. 
Setting $\eta_1 = \eta_2$ and $\bar{\eta}_1 = \bar{\eta}_2$, the 
FP-ghost propagator may be written, in $2 \times 2$ matrix notation, 
as 
%%%%%%%%%%%%%%%%%%%%%%%%%
\begin{eqnarray} 
\hat{\Delta}_g (x, y) & = & \Delta_g (x, y) \hat{\tau}_3 \, , 
\nonumber \\ 
\Delta_g (x, y) & = & - i \int \frac{d^{\, 4} P}{(2 \pi)^4} e^{- i P 
\cdot (x - y)} \frac{1}{p^2} \, . 
\label{kore12} 
\end{eqnarray} 
%%%%%%%%%%%%%%%%%%%%%%%%%%%%%%%%%%%%%%%%%%%%%%
As will be shown in Appendix A, the FP-ghost self-energy-part matrix 
$\hat{\Pi}_g$ is also diagonal, and enjoys a property $[\Pi_g]_{1 
1} + [\Pi_g]_{2 2} = 0$ with $\mbox{Im} [\Pi_g (X; P)]_{1 1} = 
0$. Then we may write 
%%%%%%%%%%%%%%%%%%%%%%%%%
\[ 
\hat{\Pi}_g (x, y) = \Pi_g (x, y) \hat{\tau}_3 \, . 
\] 
%%%%%%%%%%%%%%%%%%%%%%%%%%%%%%%%%%%%%%%%%%%%%%

A $\hat{\Pi}_g$-resummed FP-ghost propagator $\hat{G}_g$ $(= G_g 
\hat{\tau}_3)$ is defined through a Schwinger-Dyson equation: 
%%%%%%%%%%%%%%%%%%%%%%%%%
\[ 
G_g = \Delta_g + \Delta_g \cdot \Pi_g \cdot G_g 
= \Delta_g + G_g \cdot \Pi_g \cdot \Delta_g \, . 
\] 
%%%%%%%%%%%%%%%%%%%%%%%%%%%%%%%%%%%%%%%%%%%%%%
Here we have used the short-hand notation $F \cdot G$, which is a 
function whose \lq\lq $(x \, y)$-component'' is 
%%%%%%%%%%%%%%%%%%%%%%%
\begin{equation} 
[F \cdot G] (x, y) = \int d^{\, 4} z \, F (x, z) G (z, y) \, . 
\label{short} 
\end{equation} 
%%%%%%%%%%%%%%%%%%%%%%
Solving this to the gradient approximation, we obtain 
%%%%%%%%%%%%%%%%%%%%%%%%%
\begin{eqnarray*} 
G_g (X; P) & = & G_g^{(0)} (X; P) + G_g^{(1)} (X; P) \, , \nonumber 
\\ 
G_g^{(0)} (X; P) & = & - \frac{1}{p^2 + \Pi_g (X; P)} \, , \nonumber 
\\ 
G_g^{(1)} (X; P) & = & - \frac{i}{2} \frac{\partial \Pi_g (X; 
P)}{\partial P^\mu} \frac{\partial \Pi_g (X; P)}{\partial X_\mu} [ 
G^{(0)}_g (X; P) ]^3 \, . 
\end{eqnarray*} 
%%%%%%%%%%%%%%%%%%%%%%%%%%%%%%%%%%%%%%%%%%%%%%
As will be observed in Sec.~IV, to be consistent with the 
approximation we are taking, $G_g^{(1)}$ may be ignored, so that 
%%%%%%%%%%%%%%%%%%%%%%%%%
\[ 
\hat{G}_g (X; P) \simeq - \frac{\hat{\tau}_3}{p^2 + \Pi_g (X; P)} 
\, . 
\] 
%%%%%%%%%%%%%%%%%%%%%%%%%%%%%%%%%%%%%%%%%%%%%%
%%%%%%%%%%%%%%%%%%%%%%%%%%%%%%%%%%%%%%%%%%%%%%
\subsubsection{Vertex factors} 
The vertex factors may be read off from Eq.~(\ref{hat-L}): 
There is no vertex that mixes type-1 fields with type-2 fields. 
Every vertex factor for interactions between the type-1 fields is 
the same as in vacuum theory, while the every vertex factor for the 
type-2 fields is of opposite sign to the corresponding vertex factor 
for the type-1 fields. 

In the next subsection, we deduce the gluon propagator, 
Eq.~(\ref{kore}). 
%%%%%%%%%%%%%%%%%%%%%%%%%%%%%%%%%%%%%%%%%%%%%%%%%%%%
\subsection{Gluon propagator} 
We compute Eq.~(\ref{kore}) with $A_1^\mu = A^\mu_2$ 
$(\equiv A^\mu)$. For $A^\mu$, we use the plane-wave decomposition 
with helicity basis in vacuum theory. Straightforward but lengthy 
calculation yields 
%%%%%%%%%%%%%%%%%%%%%%%%%%%%%%%%%%%%%%%%%%%%%%%%%%%%%%%%%%%%%
\begin{eqnarray} 
\hat{\Delta}^{\mu \nu} (x, y) & = & \int \frac{d^{\, 4} P}{(2 
\pi)^4} \, e^{- i P \cdot (x - y)} \, \hat{\Delta}^{\mu \nu} (X; P) 
\;\;\;\;\;\;\;\;\; \left( X \equiv (x + y) / 2 \right) \, , 
\label{iti} 
\end{eqnarray} 
%%%%%%%%%%%%%%%%%%%%%
where 
%%%%%%%%%%%%%%%%%%%%%%%%%
\begin{eqnarray} 
\hat{\Delta}^{\mu \nu} (X; P) & = & \hat{\Delta}_T^{\mu \nu} (X; P) 
+ \hat{\Delta}_L^{\mu \nu} (X; P) + \hat{\Delta}_1^{\mu \nu} (X; P) 
+ \hat{\Delta}_2^{\mu \nu} (X; P) + \hat{\Delta}_3^{\mu \nu} (X; P) 
\, , \nonumber \\ 
\label{prop} 
\\ 
\hat{\Delta}_T^{\mu \nu} (X; P) & = & - {\cal P}_T^{\mu \nu} 
(\hat{\bf p}) \left[ \hat{\Delta}_{R A} (P) + \left[ \Delta_R (P) - 
\Delta_A 
(P) \right] f (X; P) \hat{M}_+ \right] \, , 
\label{2.12d} 
\\ 
\hat{\Delta}_L^{\mu \nu} (X; P) & = & \left[ n^\mu n^\nu - \alpha 
\frac{1}{p^2} P^\mu P^\nu \right] \frac{1}{p^2} \hat{\tau}_3 \, , 
\label{Lon} 
\\ 
\hat{\Delta}_1^{\mu \nu} (X; P) & = & g^{\mu i} g^{\nu j} \left[ 
\frac{i \left( 1 + (\hat{p}^y)^2 \right)}{2 p \left( 1 - 
(\hat{p}^y)^2 \right)} \left( \hat{p}^i \nabla^j_{\bf X} - \hat{p}^j 
\nabla^i_{\bf X} \right) \right. \nonumber \\ 
&& \left. - \frac{i \hat{p}^y}{p \left( 1 - (\hat{p}^y)^2 \right)} 
\left( \delta^{i y} \nabla^j_\perp - \delta^{j y} \nabla^i_\perp 
\right) \right] \left[ \Delta_R (P) - \Delta_A (P) 
\right] f (X; P) \hat{M}_+ \, , \nonumber \\ 
\label{2.666} \\ 
\hat{\Delta}_2^{\mu \nu} (X; P) & = & - i g^{\mu i} g^{\nu j} 
\epsilon^{i j k} 
\hat{p}^k N_- (P) \left[ \Delta_R (P) - \Delta_A (P) 
\right] \epsilon (p^0) \hat{M}_+ \, , 
\label{2.66} \\ 
\hat{\Delta}_3^{\mu \nu} (X; P) & = & g^{\mu i} g^{\nu j} \left\{ 
\left[ 
(1 - 2 \delta^{i y} ) \delta^{i j} + \hat{p}^i \hat{p}^j - \frac{2 
\tilde{p}^i 
\tilde{p}^j}{1 - (\hat{p}^y)^2} \right] \mbox{Re} N_{L R} (P) 
\right. \nonumber \\ 
&& - \left[ \frac{\hat{p}^y \left[ \tilde{p}^i (\hat{\bf p} \times 
{\bf e}^y)^j + \tilde{p}^j (\hat{\bf p} \times {\bf e}^y)^i 
\right]}{1 - (\hat{p}^y)^2} \right. \nonumber \\ 
&& \left. \left. - (\hat{\bf p} \times {\bf e}^y)^i \delta^{j y} 
- (\hat{\bf p} \times {\bf e}^y)^j \delta^{i y} \right] \epsilon 
(p^0) \mbox{Im} N_{L R} (P) \right\} \nonumber \\ 
&& \times \left[ \Delta_R (P) - 
\Delta_A (P) \right] \epsilon (p^0) \hat{M}_+ \, , 
%&& \left. \left. - (\hat{\bf p} \times {\bf e}^y)^i \delta^{j y} 
%- (\hat{\bf p} \times {\bf e}^y)^j \delta^{i y} \right] \epsilon 
%(p^0) \mbox{Im} N_{L R} (P) \right\} \nonumber \\ 
%&& \times \left[ \Delta_R (P) - \Delta_A (P) \right] \epsilon 
%(p^0) \hat{M}_+ \, , 
\label{dai1} 
\end{eqnarray} 
%%%%%%%%%%%%%%%%%%%%%%%%%%%%%%%%%%%%%%%%%%%%%%
where the Greek letters, $\mu$ and $\nu$, take $0$~-~$3$, while the 
Latin letters, $i$ and $j$, take $1$, $2$, and $3$. To avoid 
possible confusion, we have written $(x, y,z)$ for $(1, 2, 3)$. In 
the above equations, ${\bf e}^y = (0, 1, 0)$, $\nabla^i_{\bf X} 
\equiv \partial / \partial X^i$, $\epsilon^{1 2 3} = \epsilon^{x y 
z} = 1$, and 
%%%%%%%%%%%%%%%%%%%%%%%%%%%%%%%%%%%%%%%%%%%%%%%%%%%%%%%%%%%%%
\begin{eqnarray} 
{\cal P}^{\mu \nu}_T (\hat{\bf p}) & = & - g^{\mu i} g^{\nu j} 
\left( \delta^{i j} - \hat{p}^i \hat{p}^j \right) \;\;\;\;\;\; 
(\hat{\bf p} \equiv {\bf p} / p) \, , \\ 
\tilde{\bf p} & \equiv & \hat{\bf p} - (\hat{\bf p} \cdot {\bf e}^y) 
{\bf e}^y \, ,\nonumber \\ 
\nabla^i_\perp & \equiv & \nabla^i_{\bf X} - (\nabla_{\bf X} \cdot 
\hat{p}) \hat{p}^i \, , \nonumber \\ 
\hat{\Delta}_{R A} (P) & = & \left( 
\begin{array}{cc} 
\Delta_R (P) & \;\; 0 \\ 
\Delta_R (P) - \Delta_A (P) & \;\; - \Delta_A (P) 
\end{array} 
\right) \, , 
\label{RAM} 
\;\;\;\;\;\;\;\;\;\;\; 
\Delta_{R (A)} (P) = \frac{1}{P^2 \pm i p_0 0^+} \, , \nonumber 
\\ 
\hat{M}_{\pm} & = & \left( 
\begin{array}{cc} 
1 & \;\; \pm1 \\ 
\pm 1 & 1 
\end{array} 
\right) \, , 
\label{Apm} \\ 
f (X; P) & = & \epsilon (p_0) N (X; P) - \theta (- p_0) \, , 
\label{f-desu} 
\\ 
N (X; P) & \equiv & \frac{1}{2} \left[ N_{+ +} (X; |p^0|, \epsilon 
(p^0) \hat{\bf p}) +  N_{- -} (X; |p^0|, \epsilon (p^0) \hat{\bf p}) 
\right] \nonumber \\ 
& \equiv & \theta (p^0) n (X; |p^0|, \hat{\bf p}) + \theta (- p^0) 
n (X; |p^0|, - \hat{\bf p}) \, , 
\label{N-desu} 
\\ 
N_- (X; P) & \equiv & \frac{1}{2} \left[ N_{+ +} (X; |p^0|, \epsilon 
(p^0) \hat{\bf p}) - N_{- -} (X; |p^0|, \epsilon (p^0) \hat{\bf p}) 
\right] \nonumber \\ 
& \simeq & N_- (P) \, , 
\label{2.23d} \\ 
N_{L R} (P) & \equiv & N_{- +} (|p_0|, \epsilon (p^0) \hat{\bf p}) 
\label{yatto} 
\end{eqnarray} 
%%%%%%%%%%%%%%%%%%%%%%%%%%%%%%%%%%%%%%%%%%%%%%
with 
%%%%%%%%%%%%%%%%%%%%%%%%%%%%%%%%%%%%%%%%%%%%%%%%%%%%%%%%%%%%%
\begin{eqnarray} 
N_{\xi \zeta} (X; |p_0|, \hat{\bf p}) & \equiv & \int d^{\, 3} q 
\, e^{- i Q \cdot X} \mbox{Tr} \left[ 
a^\dagger_\xi ({\bf p} - {\bf q} / 2) \, a_\zeta ({\bf p} + {\bf 
q} / 2) \, \rho \right] 
\;\;\;\;\;\; (\xi, \zeta = +, -) \, . \nonumber \\ 
\label{no} 
\end{eqnarray} 
%%%%%%%%%%%%%%%%%%%%%%%%%%%%%%%%%%%%%%%%%%%%%%
Here $|p_0| = p$, and $Q = (q^0, {\bf q})$ with $q_0 = 
{\bf q} \cdot {\bf p} / p$. $a_+$ ($a_-$) is an annihilation 
operator of a gluon with $+$ ($-$) helicity. $a_+^\dagger$ 
($a_-^\dagger$) is a corresponding creation operator. In deriving 
Eqs.~(\ref{prop})~-~(\ref{no}), we have assumed that $\mbox{Tr} 
[a_\rho ({\bf p}) a_\sigma ({\bf q}) \, \rho] \simeq 0$. We have 
also assumed that $N_-$, Eq.~(\ref{2.23d}), and $N_{L R}$, 
Eq.~(\ref{yatto}), are small quantities, so that 
$\hat{\Delta}_2^{\mu \nu}$ and $\hat{\Delta}_3^{\mu \nu}$ are of the 
order of gradient term $\hat{\Delta}_1^{\mu \nu}$ that includes 
$\nabla_{\bf X} N$. More precisely $N_-$ and $N_{L R} (P)$ are of 
order of $\left( \nabla_{\bf X} N \right) / p$, and then, to the 
gradient approximation, $X$-dependences of $N_-$ and of $N_{L R}$ 
have been ignored. 

We observe that the structure of $\hat{\Delta}^{\mu \nu}_L (P)$ in 
Eq.~(\ref{Lon}) is rather simple, which represents the propagator 
between \lq\lq longitudinal gluons'' as well as the gauge part. This 
is a reflection of the fact that, in the Coulomb gauge, only 
transverse modes are physical modes. Then, as in the FP-ghost 
propagator in Eq.~(\ref{kore12}), the $(1 \, 1)$-component of 
$\hat{\Delta}_L^{\mu \nu}$, $[\hat{\Delta}_L^{\mu \nu} (P)]_{1 1}$, 
is the same as in vacuum theory. 

From Eq.~(\ref{no}), follows 
%%%%%%%%%%%%%%%%%%%%%%%
\begin{equation} 
\theta (\pm p^0) P \cdot \partial_X N_{\xi \zeta} (X; |p^0|, \pm 
\hat{\bf p}) = 0 \;\;\;\;\;\; (|p_0| = p) \, . 
\label{ba} 
\end{equation} 
%%%%%%%%%%%%%%%%%%%%%%%%%%
Using this in Eqs.~(\ref{f-desu}) and (\ref{N-desu}), we 
have 
%%%%%%%%%%%%%%%%%%%%%%%
\begin{equation} 
P \cdot \partial_X f (X; P) = 0 \;\;\;\;\;\; (|p_0| = p) \, . 
\label{ba1} 
\end{equation} 
%%%%%%%%%%%%%%%%%%%%%%%%%%
As is obvious from the construction, or as can be directly shown 
from Eq.~(\ref{ba1}), we see that 
%%%%%%%%%%%%%%%%%%%%%%%%%%%%%%%%%%%%%%%%%%%%%%%%%%%%%%%%%%%%%%%
\begin{equation} 
- \hat{\tau}_3 {\cal D}^{\mu \rho} \hat{\Delta}_{\rho \nu} 
(x, y) = - \hat{\Delta}^{\mu \rho} (x, y) {\cal D}_{\rho \nu} 
\hat{\tau}_3 = \delta^\mu_{\;\; \nu} \delta^{\, 4} (x - y) \, , 
\label{2.26d} 
\end{equation} 
%%%%%%%%%%%%%%%%%%%%%%%%%%%%%%%%%%%%%%%%%%%%%%%%
where ${\cal D}^{\mu \rho}$ is as in Eq.~(\ref{Ddesu}). Two 
equations in Eq. (\ref{ba}) are \lq\lq free Boltzmann 
equations.'' One can construct a perturbation theory in a similar 
manner as in \cite{nie}, where a complex-scalar field system is 
treated. We call the perturbation 
theory thus constructed the bare-$N$ scheme, since $N$ obeys the 
\lq\lq free Boltzmann equation.'' This scheme is equivalent 
\cite{nie} to the physical-$N$ scheme, to which we now turn. 
%%%%%%%%%%%%%%%%%%%%%%%%%%%%%%%%%%%%%%%%%%%%%%
%%%% SECTION III %%%%%%%%%%%%%%%%%%%%%%%%%%%%%%%
%%%%%%%%%%%%%%%%%%%%%%%%%%%%%%%%%%%%%%%%%%%%%%
\section{Construction of the physical-$N$ scheme} 
We aim to construct a scheme in terms of the number density that is 
as close as possible to the physical number density. To this end, 
first of all, we abandon the \lq\lq free Boltzmann equation'' 
(\ref{ba1}). This means that $f$ and $A$ in the present 
(physical-$N$) scheme differs from the bare-$N$ scheme 
counterparts. Specification of $f$ is postponed until Sec.~VI. 

Now, in contrast to Eq.~(\ref{2.26d}), $\hat{\Delta}^{\mu \nu}$ is 
not an inverse of $- \hat{\tau}_3 {\cal D}_{\mu \nu}$. 
Straightforward calculation within the gradient approximation yields 
%%%%%%%%%%%%%%%%%%%%%%%%%%%%%%%%
\begin{eqnarray*} 
\hat{\tau}_3 {\cal D}^\mu_{\; \, \rho} \, \hat{\Delta}^{\rho \nu} 
(x, y) & = & - g^{\mu \nu} \delta^{\, 4} (x - y) 
- i \hat{\tau}_3 \hat{M}_+ \int \frac{d^{\, 4} P}{(2 \pi)^4} 
e^{- i P \cdot (x - y)} \nonumber \\ 
&& \times \left[ \Delta_R (P) - \Delta_A (P) 
\right] {\cal P}_T^{\mu \nu} (\hat{\bf p}) \left( P \cdot 
\partial_X f (X; P) 
\right) \, . 
\end{eqnarray*} 
%%%%%%%%%%%%%%%%%%%%%%%%%%%%%%%%%%%%%%%%%%%%%
In obtaining this, we have used $P^2 [ \Delta_R (P^2) - \Delta_A 
(P^2) ] \propto P^2 \delta(P^2) = 0$. Our procedure of constructing 
a consistent scheme is as follows: We further modify $A^\mu$ by 
adding a suitable $\hat{\Delta}_{\mbox{\scriptsize{add}}}^{\rho 
\nu}$ to $\hat{\Delta}^{\rho \nu}$. The conditions for 
$\hat{\Delta}_{\mbox{\scriptsize{add}}}^{\rho \nu}$ to be satisfied 
are 
\begin{itemize} 
\item $\hat{\Delta}_{\mbox{\scriptsize{add}}}^{\rho \nu}$ vanishes 
in the bare-$N$ scheme. 
\item To the gradient approximation, the counterpart of 
Eq.~(\ref{2.26d}) turns out to 
%%%%%%%%%%%%%%%%%%%%%%%%%%%%%%%%%%%%%%%%%%%%%
\begin{equation} 
\left( - \hat{\tau}_3 {\cal D}_{\mu \rho} 1 - \left( \hat{L}_c 
\right)_{\mu \rho} \right) \cdot \left( 
\hat{\Delta}^{\rho \nu} + 
\hat{\Delta}_{\mbox{\scriptsize{add}}}^{\rho \nu} \right) = 
\delta_\mu^{\;\; \nu} 1 \, , 
\label{ma} 
\end{equation} 
%%%%%%%%%%%%%%%%%%%%%%%%%%%%%%%%
where use has been made of the short-hand notation (\ref{short}). 
\lq $1$' is the matrix in spacetime whose $(x \, y)$-component is 
$\delta^{\, 4} (x - y)$, and $\hat{L_c}$ is some $(4 \times 4) 
\otimes (2 \times 2)$ matrix function. 
\end{itemize} 

It is a straightforward task to obtain the form of the required 
$\hat{\Delta}_{\mbox{\scriptsize{add}}}^{\mu \nu}$: 
%%%%%%%%%%%%%%%%%%%%%%%%%%%%%%%%
\begin{equation} 
\hat{\Delta}_{\mbox{\scriptsize{add}}}^{\mu \nu} (X;. P) = i 
\hat{M}_+ {\cal P}_T^{\mu \nu} ( \hat{\bf p}) \left[ \Delta_R^2 (P) 
+ \Delta_A^2 (P) \right] P \cdot \partial_X f (X; P) \, , 
\label{add} 
\end{equation} 
%%%%%%%%%%%%%%%%%%%%%%%%%%%%%%%%%%%%%%%%%%%%%
from which we obtain for $\hat{L}_c^{\mu \nu}$ in Eq.~(\ref{ma}), 
%%%%%%%%%%%%%%%%%%%%%%%%%%%%%%%%
\begin{eqnarray*} 
\hat{L}_c^{\mu \nu} (x, y) & = & L_c^{\mu \nu} \hat{M}_- \nonumber 
\\ 
& = & 2 i \hat{M}_- \int \frac{d^{\, 4} P}{(2 \pi)^4} e^{- i P \cdot 
(x - y)} \, {\cal P}_T^{\mu \nu} (\hat{\bf p}) P \cdot \partial_X f 
(X; P) \, , 
\end{eqnarray*} 
%%%%%%%%%%%%%%%%%%%%%%%%%%%%%%%%%%%%%%%%%%%%%
with $\hat{M}_-$ as in Eq.~(\ref{Apm}). In obtaining Eq.~(\ref{ma}) 
with $\hat{\Delta}_{\mbox{\scriptsize{add}}}^{\mu \nu}$ as in 
Eq.~(\ref{add}), we have used $(\hat{L}_c)^\mu_{\;\; \nu} \cdot 
\hat{\Delta}_0^{\nu \rho} \simeq (\hat{L}_c)^\mu_{\; \; \nu} 
\cdot (\hat{\Delta}^{\nu \rho} + 
\hat{\Delta}_{\mbox{\scriptsize{add}}}^{\nu \rho})$, since the 
difference can be ignored to the gradient approximation.  
In a similar manner, we find that $(\hat{\Delta}^{\mu \rho} + 
\hat{\Delta}_{\mbox{\scriptsize{add}}}^{\mu \rho}) \cdot ( - 
\hat{\tau}_3 1 \stackrel{\leftarrow}{\cal D}_{\rho \nu} - ( 
\hat{L}_c )_{\rho \nu}) = \delta^\mu_{\;\; \nu} 1$. Thus we have 
found that the $2 \times 2$ matrix propagator $(\hat{\Delta}^{\mu 
\nu} + \hat{\Delta}_{\mbox{\scriptsize{add}}}^{\mu \nu})$ is an 
inverse of $(- \hat{\tau}_3 {\cal D}_{\mu \nu} 1 - ( \hat{L}_c 
)_{\mu \nu})$, so that the free action is\footnote{Note that the 
Lagrangian density corresponding to the term with $L_c^{\mu \nu} (x, 
y)$ in Eq.~(\ref{acti}) is nonlocal not only in \lq space' but also 
in \lq time.' Here it is worth mentioning the so-called 
$|p_0|$-prescription. With this prescription, at an intermediate 
stage, we have $\left( L_c' \right)^{\mu \nu} (x, y)$, which is 
local in time. [See \cite{nie,nie1} for details.]}   
%%%%%%%%%%%%%%%%%%%%%%%%%%%%%%%%%%%%%%%%%%%%%%%%%%%%%%%%%%%%%
\begin{eqnarray} 
&& - \frac{1}{2} \int d^{\, 4} x \, d^{\, 4} y \, {}^t \! 
\hat{A}_\mu (x) \left[ \hat{\tau}_3 {\cal D}^{\mu \nu} \delta^{\, 4} 
(x - y) + L_c^{\mu \nu} (x, y) \hat{M}_- 
\right] \hat{A}_\nu (y) \, , \nonumber \\ 
&& \mbox{\hspace*{25ex}} \;\;\;\;\; \hat{A}^\mu = \left( A_1^\mu, 
\, A_2^\mu \right) \, . 
\label{acti} 
\end{eqnarray} 
%%%%%%%%%%%%%%%%%%%%%%%%%%%%%%%%%%%%%%%%%%%%%%%%%%%%%%%
Since the term with $L_c (x, y)$ in Eq.~(\ref{acti}) is absent in 
the original action, we should introduce the counter action to 
compensate it, 
%%%%%%%%%%%%%%%%%%%%%%%%%%%%%%%%%%%%%%%%%%%%
\begin{equation} 
{\cal A}_c = \frac{1}{2} \int d^{\, 4} x \, d^{\, 4} y \, {}^t \! 
\hat{A}_\mu (x) L_c^{\mu \nu}  (x, y) \hat{M}_- \hat{A}_\nu (y) \, , 
\label{act} 
\end{equation} 
%%%%%%%%%%%%%%%%%%%%
which yields a vertex 
%%%%%%%%%%%%%%%%%%%%%%%%%%%%%%%%%%%%%%%%%%%%
\begin{eqnarray} 
i L_c^{\mu \nu} (x, y) \hat{M}_- & = & - 2 \hat{M}_- \int 
\frac{d^{\, 4} P}{(2 \pi)^4} e^{- i P \cdot (x - y)} {\cal P}^{\mu 
\nu}_T (\hat{\bf p}) P \cdot \partial_X f (X; P) 
\nonumber \\ 
& \equiv & i \hat{\Pi}^{(c) \mu \nu} (x, y) \, . 
\label{sigmac} 
\end{eqnarray} 
%%%%%%%%%%%%%%%%%%%%

For completeness, we construct a $\hat{L}_c$-resummed propagator. 
Since $\hat{M}_+ \hat{M}_- = \hat{M}_- \hat{M}_+ = 0$ and 
$\hat{\Pi}^{(c) \mu \nu} \propto {\cal P}_T^{\mu \nu}$, only piece 
that changes the form under the $\hat{L}_c$-resummation is 
$- {\cal P}_T^{\mu \nu} (\hat{\bf p}) \hat{\Delta}_{R A}$ in 
$\hat{\Delta}_T^{\mu \nu}$, Eq.~(\ref{2.12d}) with Eq.~(\ref{RAM}): 
%%%%%%%%%%%%%%%%%%%%%%%%%%%%%%%%%%%%%%%%%%%%
\begin{eqnarray*} 
- {\cal P}_T^{\mu \nu} (\hat{\bf p}) \hat{\Delta}_{R 
A}^{\mbox{\scriptsize{resum}}} (x, y) & = & - {\cal P}_T^{\mu \nu} 
(\hat{\bf p}) \left\{ \hat{\Delta}_{R A} (x, y) - \sum_{n = 
1}^\infty \left[ \hat{\Delta}_{R A} \left\{ \cdot \hat{\Pi}^{(c)} 
\cdot \hat{\Delta}_{R A} \right\}^n \right] (x, y) \right\} 
\nonumber \\ 
& = & - {\cal P}_T^{\mu \nu} (\hat{\bf p}) \left\{ \hat{\Delta}_{R 
A}^{\mu \nu} (x, y) - \left[ \hat{\Delta}_{R A} \cdot 
\hat{\Pi}^{(c)} \cdot \hat{\Delta}_{R A} \right] (x, y) \right\} 
\nonumber \\ 
& \simeq & - {\cal P}_T^{\mu \nu} (\hat{\bf p}) \left[ 
\hat{\Delta}_{R A} (x, y) + 2 i \hat{M}_+ \int \frac{d^{\, 4} 
P}{(2 \pi)^4} e^{- i P \cdot (x - y)} \right. \nonumber \\ 
&& \left. \times \Delta_R (P) \Delta_A (P) \, P \cdot \partial_X f 
(X; P) \right] \, . 
\end{eqnarray*} 
%%%%%%%%%%%%%%%%%%%%
Here, use has been made of $\hat{\Delta}_{R A}^{\mu \nu} \hat{M}_- 
\hat{\Delta}_{R A}^{\mu \nu} \propto \hat{M}_+$. Then, substituting 
$\left( \hat{\Delta}_{R A}^{\mbox{\scriptsize{resum}}} \right)^{\mu 
\nu}$ for $\hat{\Delta}_{R A}^{\mu \nu}$, we obtain, with obvious 
notation, 
%%%%%%%%%%%%%%%%%%%%%%%%%%%%%%%%%%%%%%%%%%%%
\begin{eqnarray} 
&& \left( \hat{\Delta}_T^{\mbox{\scriptsize{resum}}} (X; P) 
\right)^{\mu \nu} + \hat{\Delta}_{\mbox{\scriptsize{add}}}^{\mu \nu} 
(X; P) \nonumber \\ 
& & \mbox{\hspace*{5ex}} = - {\cal P}_T^{\mu \nu} (\hat{\bf p}) 
\left\{ \left[ \hat{\Delta}_{R A} (X; P) + \hat{M}_+ \left[ \Delta_R 
(P) - \Delta_A (P) \right] f (X; P) \right] 
\right. \nonumber \\ 
&& \mbox{\hspace*{7.8ex}} \left. - i \hat{M}_+ {\cal P}_T^{\mu \nu} 
(\hat{\bf p}) \left[ \Delta_R (P) - \Delta_A (P) \right]^2 P \cdot 
\partial_X f (X; P) \right\} \, . 
\label{ressp} 
\end{eqnarray} 
%%%%%%%%%%%%%%%%%%%%
In closing this section, we emphasize that $f (X; P)$ in the present 
scheme is an arbitrary function, provided that $f (X^0_{in}, {\bf 
X}; P)$ [$= - \theta (- p^0) + \epsilon (p^0) N(X^0_{in}, {\bf 
X}; P)$], with $X_0^{in}$ the initial time, is a given initial data. 
We have introduced the counteraction ${\cal A}_c$, Eq.~(\ref{act}), 
so as to remain on the original theory. Thus, it cannot be 
overemphasized that the schemes with different $f$'s are mutually 
equivalent. If we choose the $f$ $(= f_B)$ that subjects to the 
\lq\lq free Boltzmann equation'' (\ref{ba1}), the scheme reduces to 
the bare-$N$ scheme in the last section. In Sec.~VI, we shall 
choose $f$, with which a well-defined perturbation theory is 
formulated. In any case, it is natural to choose $f$, so that, when 
interactions are switched off, $f$ turns out to $f_B$. 
%%%%%%%%%%%%%%%%%%%%%%%%%%%%%%%%%%%%%%%%%%%%%%
%%%% SECTION III %%%%%%%%%%%%%%%%%%%%%%%%%%%%%%%
%%%%%%%%%%%%%%%%%%%%%%%%%%%%%%%%%%%%%%%%%%%%%%
\section{Resummation of the self-energy part} 
%%%%%%%%%%%%%%%%%%%%%%%%%%%%%%%%%%%%%%%%%%%%
\subsection{Preliminary} 
Throughout in the sequel, we restrict to the strict Coulomb gauge 
$\alpha = 0$ [cf. Eq.~(\ref{Ddesu})]. The bare propagator consists 
of six pieces, $\hat{\bf 
\Delta}_T + \hat{\bf \Delta}_L + \sum_{i = 1}^3 \hat{\bf \Delta}_i 
+ \hat{\bf \Delta}_{\mbox{\scriptsize{add}}}$, Eqs.~(\ref{prop}) 
and (\ref{add}). $\hat{\bf \Delta}_T + \hat{\bf \Delta}_L$ is the 
leading part and $\hat{\bf \Delta}_1$ and $\hat{\bf 
\Delta}_{\mbox{\scriptsize{add}}}$ are the gradient parts, which 
represents variation in the macroscopic spacetime coordinates 
$X_\mu$, through first-order derivative $\partial_{X_\mu} f (X; P)$. 
We have assumed that $\hat{\bf \Delta}_2$ and $\hat{\bf \Delta}_3$ 
are of the same order of magnitude as the gradient parts. 
Interactions among the fields give rise to reactions taking place in 
a system, which, in turn, causes a nontrivial change in the number 
density of quasiparticles. Thus, the self-energy part $\hat{\bf 
\Pi}$ ties with the gradient parts. More precisely, $\hat{\bf 
\Pi}$ is of the same order of magnitude as $\hat{\bf 
\Delta}_{\mbox{\scriptsize{add}}} \left( \hat{\bf \Delta}_T + 
\hat{\bf \Delta}_L \right)^{- 2}$ and $\hat{\bf \Delta}_i \left( 
\hat{\bf \Delta}_T + \hat{\bf \Delta}_L \right)^{- 2}$ $(i = 1 - 
3)$. Hence, in computing $\hat{\bf \Pi}$ in the approximation under 
consideration, it is sufficient to keep the leading part (i.e., the 
part with no $X_\mu$-derivative). Then $\hat{\bf \Pi} (X; P)$ may be 
decomposed as 
%%%%%%%%%%%%%%%%%%%%%%%%%%%%%%%%%%%%%%%%%%%%
\begin{eqnarray} 
\hat{\Pi}^{\mu \nu} (X; P) & = & {\cal P}_T^{\mu \nu} (\hat{\bf p}) 
\hat{\Pi}_T (X; P) + n^\mu n^\nu \hat{\Pi}_L (X; P) \nonumber \\ 
&& + \frac{p_0}{p} \left( n^\mu \hat{p}^{\underline{\nu}} + 
\hat{p}^{\underline{\mu}} n^\nu \right) \hat{\Pi}_C (X; P) + 
\hat{p}^{\underline{\mu}} \hat{p}^{\underline{\nu}} \hat{\Pi}_D (X; 
P) \, . 
\label{koo} 
\end{eqnarray} 
%%%%%%%%%%%%%%%%%%%%%%%%%%%%%%%%%%%%%%%%%%%%
Here and in the following, we use such a notation as 
$v^{\underline{\mu}}$ for denoting a four-vector whose $0$th 
component is zero: $v^{\underline{\mu}} \equiv (0, {\bf v})$. 
Then, $\hat{p}^{\underline{\mu}} = (0, \hat{\bf p})$.

Within the gradient approximation, it is sufficient to perform a 
$\hat{\bf \Pi}$-resummation for the leading part, $\hat{\bf 
\Delta}_T + \hat{\bf \Delta}_L$. This is because the corrections to 
other pieces due to the resummation are of higher order. Thus, for 
the gradient parts of $\hat{\bf \Delta}$ as well as for $\hat{\bf 
\Delta}_{\mbox{\scriptsize{add}}}$, one can use the formulae in the 
bare-$N$ scheme in Sec.~II [cf. above argument after 
Eq.~(\ref{ressp})]. In particular, for $f$ in the gradient parts, 
one can use $f$ ($ = f_B$) as in Eq.~(\ref{ba1}) in the bare-$N$ 
scheme. For the purpose of performing resummation, as usual, it is 
convenient to introduce the \lq\lq standard form'' 
\cite{ume,nie,nie1,nie2}
%%%%%%%%%%%%%%%%%%%%%%%%%%%%%%%%%%%%%%%%%%%%
\begin{eqnarray} 
\left[ \hat{B}_L \cdot \hat{\bf 
\Delta}_{\mbox{\scriptsize{diag}}} \cdot \hat{B}_R \right] (x, y) 
& = & \int d^{\, 4} z_1 \int d^{\, 4} z_2 \, 
\hat{B}_L (x, z_1) \hat{\bf \Delta}_{\mbox{\scriptsize{diag}}} (z_1, 
z_2) \hat{B}_R (z_2, y) \, , 
\label{sta} \\ 
\hat{\Delta}_{\mbox{\scriptsize{diag}}}^{\mu \nu} & = & \mbox{diag} 
\left( \Delta_R^{\mu \nu} , \; - \Delta_A^{\mu \nu} \right) \, , 
\nonumber \\ 
\Delta_{R (A)}^{\mu \nu} (P) & = & - \tilde{g}^{\mu \nu} (P) 
\Delta_{R (A)} (P) \, , \nonumber \\ 
\hat{B}_L & = & \left( 
\begin{array}{cc} 
1 & \;\; f \\ 
1 & \;\; 1 + f 
\end{array} 
\right) \, , \;\;\;\;\; 
\hat{B}_R = \left( 
\begin{array}{cc} 
1 + f & \;\; f \\ 
1 & \;\; 1 
\end{array} 
\right) \, , 
\nonumber \\ 
\tilde{g}^{\mu \nu} (P) & = & g^{\mu \nu} + \frac{1}{p^2} \left( 1 + 
\alpha \frac{P^2}{p^2} \right) P^\mu P^\nu - \frac{p^0}{p^2} \left( 
P^\mu n^\nu + P^\nu n^\mu \right) \nonumber \\ 
& = & {\cal P}_T^{\mu \nu} ({\bf p}) - 
\frac{P^2}{p^2} n^\mu n^\nu + \alpha \frac{P^2}{p^4} P^\mu P^\nu \, 
, 
\label{4.4d} 
\end{eqnarray} 
%%%%%%%%%%%%%%%%%%%%%%%%%%%%%%%%%%%%%%%%%%%%
where $f$ $(= f (x, y))$ is the inverse Fourier transform of $f (X; 
P)$. It is to be noted that $- \Delta^{\mu \nu}_{R (A)} (x - y)$ is 
an inverse of ${\cal D}_\mu^{\;\, \nu}$, Eq.~(\ref{Ddesu}): 
%%%%%%%%%%%%%%%%%%%%%%%%%%%%%%%%%%%%%%%%%%%%
\begin{equation} 
{\cal D}^\mu_{\;\, \rho} \Delta^{\rho \nu}_{R (A)} = 
\Delta^{\mu \rho}_{R (A)} \stackrel{\leftarrow}{{\cal D}_{\;\, 
\rho}^\nu} = - g^{\mu \nu} \, . 
\label{4.5d} 
\end{equation} 
%%%%%%%%%%%%%%%%%%%%%%%%%%%%%%%%%%%%%%%%%%%%
Computing Eq.~(\ref{sta}) to the gradient approximation, we obtain 
%%%%%%%%%%%%%%%%%%%%%%%%%%%%%%%%%%%%%%%%%%%%
\begin{eqnarray} 
\hat{\Delta}^{\mu \nu}_{\scriptsize{\mbox{base}}} (x, y) & \equiv & 
\hat{\Delta}_T^{\mu \nu} (x, y) 
+ \hat{\Delta}_L^{\mu \nu} (x, y) 
+ \hat{\Delta}^{\mu 
\nu}_{\mbox{\scriptsize{add}}} (x, y) \nonumber \\ 
& = & \left[ \hat{B}_L \cdot 
\hat{\Delta}_{\scriptsize{\mbox{diag}}}^{\mu \nu} \cdot \hat{B}_R 
\right] (x, y) + \hat{M}_+ \Delta_K^{\mu \nu} (x, y) \, , 
\label{4.555} 
\\ 
\Delta_K^{\mu \nu} (X; P) & = & - \frac{i}{p^2} \left[ \frac{\bf 
P}{P^2} \left( p^{\underline{\mu}} \nabla_\perp^{\underline{\nu}} + 
p^{\underline{\nu}} \nabla_\perp^{\underline{\mu}} \right) f + 2 
\frac{n^\mu n^\nu}{p^2} {\bf p} \cdot {\bf \nabla_{\bf X}} f \right] 
\, , 
\label{itt} 
\end{eqnarray} 
%%%%%%%%%%%%%%%%%%%%%%%%%%%%%%%%%%%%%%%%%%%%%%%
where ${\bf P} / P^2$ is the principal part of $1 / (P^2 \pm i 
0^+)$. As mentioned above, within the gradient approximation, it is 
sufficient to take $\hat{\bf \Delta}_T + \hat{\bf \Delta}_L$ as a 
\lq\lq resummed part.'' It is obvious from the above observation 
that one can freely include gradient part(s) into the 
\lq\lq resummed part.'' We include the gradient part 
$\hat{\Delta}^{\mu 
\nu}_{\mbox{\scriptsize{add}}}$ and take 
$\hat{\Delta}^{\mu 
\nu}_{\mbox{\scriptsize{base}}}$ as the 
\lq\lq resummed part.'' 

It is to be noted that, from Eqs.~(\ref{2.12d}), (\ref{Lon}), 
(\ref{RAM}), and (\ref{add}), follows 
%%%%%%%%%%%%%%%%%%%%%%%%%%%%%%%%%%%%%%%%%%%%
\begin{equation} 
\sum_{i, \, j = 1}^2 (-)^{i + j} ({\bf 
\Delta}_{\scriptsize{\mbox{base}}})_{i j} = 0 \, . 
\label{meme} 
\end{equation} 
%%%%%%%%%%%%%%%%%%%%%%%%%%%%%%%%%%%%%%%%%%%%
This relation holds for $\hat{\bf \Delta} + 
\hat{\bf \Delta}_{\mbox{\scriptsize{add}}}$ (in place of 
$\hat{\bf \Delta}_{\scriptsize{\mbox{base}}}$). 
%%%%%%%%%%%%%%%%%%%%%%%%%%%%%%%%%%%%%%%%%%%%%%%
%%%%%% SUB SEC %%%%%%%%%%%%%%%%%%%%%%%%%%%%%%%%%%%%%%%%%%%%%%%%%%
%%%%%%%%%%%%%%%%%%%%%%%%%%%%%%%%%%%%%%
\subsection{Self-energy-part resummed propagator} 
As will be shown in Appendix A [cf. Eq.~(\ref{jiko0})], 
%%%%%%%%%%%%%%%%%%%%%%%%%%%%
\begin{equation} 
\sum_{i, \, j = 1}^2 {\bf \Pi}_{i j} = 0 
\label{4.66} 
\end{equation} 
%%%%%%%%%%%%%%%%%%%%%%%%%%%%
holds. A $\hat{\bf \Pi}$-resummed propagator $\hat{\bf G}$ obeys 
the Sch\-win\-ger-Dyson equation: 
%%%%%%%%%%%%%%%%%%%%%%%%%%%%%%%%%%%%%%%%%%%%%%%%%%%%%%%%%%%%%%%%
\begin{equation} 
\hat{\bf G} = \hat{\bf \Delta}_{\scriptsize{\mbox{base}}} - \hat{\bf 
\Delta}_{\scriptsize{\mbox{base}}} \cdot \hat{\bf \Pi} \cdot 
\hat{\bf G} = \hat{\bf \Delta}_{\scriptsize{\mbox{base}}} - \hat{\bf 
G} \cdot \hat{\bf \Pi} \cdot \hat{\bf 
\Delta}_{\scriptsize{\mbox{base}}} \, . 
\label{SD} 
\end{equation} 
%%%%%%%%%%%%%%%%%%%%%%%%%%%%
It is clear that, for solving this equation, it is convenient to 
introduce the \lq\lq $\hat{B}$-transformed'' quantities 
%%%%%%%%%%%%%%%%%%%%%%%%%%%%%%%%%%%%%%%%%%%%%%%%%%%%%%%%%%%%%%%%
\begin{eqnarray} 
\underline{\bf \hat{G}} & \equiv & \hat{B}_L^{- 1} \cdot \hat{\bf G} 
\cdot \hat{B}_R^{- 1} \, , 
\label{hen} 
\\ 
\underline{\hat{\bf \Delta}}_{\scriptsize{\mbox{base}}} & \equiv & 
\hat{B}_L^{- 1} \cdot \hat{\bf \Delta}_{\scriptsize{\mbox{base}}} 
\cdot \hat{B}_R^{- 1} = \left( 
\begin{array}{cc} 
{\bf \Delta}_R & \;\;\, {\bf \Delta}_K \\ 
0 & \;\;\, - {\bf \Delta}_A 
\end{array} 
\right) \, , 
\label{4.11d} 
\\ 
\underline{\hat{\bf \Pi}} & \equiv & \hat{B}_R \cdot \hat{\bf \Pi} 
\cdot \hat{B}_L = \left( 
\begin{array}{cc} 
{\bf \Pi}_R & \;\;\, {\bf \Pi}_K \\ 
0 & \;\;\, - {\bf \Pi}_A 
\end{array} 
\right) \, . \nonumber 
\end{eqnarray} 
%%%%%%%%%%%%%%%%%%%%%%%%%%%%
Here ${\bf \Delta}_K$ is as in Eq.~(\ref{itt}), and 
${\bf \Pi}_{R (A)}$ and ${\bf \Pi}_K$ are, in respective order, the 
inverse Fourier transforms of 
%%%%%%%%%%%%%%%%%%%%%%%%%%%%%%%%%%%%%%%%%%%%%%%%%%%%%%%%%%
\begin{eqnarray} 
{\bf \Pi}_{R (A)} (X; P) & = & {\bf \Pi}_{1 1} + {\bf \Pi}_{1 2 (2 
1)} 
\label{SK} \\ 
{\bf \Pi}_K (X; P) & = & [ 1 + f (X; P)] {\bf \Pi}_{1 2} (X; P) 
- f (X; P) {\bf \Pi}_{2 1} (X; P) \nonumber \\ 
&& - \frac{i}{2} \left\{ {\bf \Sigma}_R + {\bf \Sigma}_A, \; f 
\right\}_{\mbox{\scriptsize{P.B.}}} \, . 
\label{SigK} 
\end{eqnarray} 
%%%%%%%%%%%%%%%%%%%%%%%%%%%%%%%%%%%%%%%%
In Eq.~(\ref{SigK}), $\{... , \, ... 
\}_{\mbox{\scriptsize{P.B.}}}$ is defined as 
%%%%%%%%%%%%%%%%%%%%%%%%%%%%%%%%%%%%%%%%%%%%%
\begin{equation} 
\left\{ A , \; B \right\}_{\mbox{\scriptsize{P.B.}}} \equiv 
\frac{\partial A (X; P)}{\partial X^\mu} \frac{\partial B (X; 
P)}{\partial P_\mu} - \frac{\partial A (X; P)}{\partial P_\mu} 
\frac{\partial B (X; P)}{\partial X^\mu} \, . 
\label{PB} 
\end{equation} 
%%%%%%%%%%%%%%%%%%%%%%%%%%%%%%%%%%%%%%%%%%%%%%%%
In obtaining Eq.~(\ref{4.11d}) [Eqs.~(\ref{SK}) and (\ref{SigK})], 
use has been made of Eq.~(\ref{meme}) [Eq.~(\ref{4.66})]. Although 
the last term in Eq.~(\ref{SigK}) may be dropped to the 
approximation under consideration, we have kept it. 

Now Eq.~(\ref{SD}) is transformed into 
%%%%%%%%%%%%%%%%%%%%%%%%%%%%%%%%%%%%%%%%%%%%%%%%%%%%%%%%%%%%%%%%
\[ 
\underline{\hat{\bf G}} = \underline{\hat{\bf 
\Delta}}_{\scriptsize{\mbox{base}}} - \underline{\hat{\bf 
\Delta}}_{\scriptsize{\mbox{base}}} \cdot \underline{\hat{\bf \Pi}} 
\cdot \underline{\hat{\bf G}} = \underline{\hat{\bf 
\Delta}}_{\scriptsize{\mbox{base}}} - \underline{\hat{\bf G}} \cdot 
\underline{\hat{\bf \Pi}} \cdot \underline{\hat{\bf 
\Delta}}_{\scriptsize{\mbox{base}}} \, , 
\] 
%%%%%%%%%%%%%%%%%%%%%%%%%%%%
which can easily be solved to give 
%%%%%%%%%%%%%%%%%%%%%%%%%%%%%%%%%%%%%%%%%%%%%
\begin{eqnarray} 
\underline{\hat{\bf G}} & = & \left( 
\begin{array}{cc} 
{\bf G}_R & \;\;\, {\bf G}_K \\ 
0 & \;\;\, - {\bf G}_A 
\end{array} 
\right) \, , 
\nonumber \\ 
{\bf G}_{R (A)} & = & \left[ {\bf \Delta}_{R (A)} + {\bf \Pi}_{R 
(A)} \right]^{- 1} \, , 
\label{3.101} 
\\ 
{\bf G}_K & = & {\bf G}_R \cdot \left[ {\bf \Delta}_R^{- 1} \cdot 
{\bf \Delta}_K \cdot {\bf \Delta}_A^{- 1} + {\bf \Pi}_K \right] 
\cdot {\bf G}_A \, . 
\label{GK} 
\end{eqnarray} 
%%%%%%%%%%%%%%%%%%%%%%%%%%%%%%%%%%%%%%%%%%%%%
Substituting this back into Eq.~(\ref{hen}), we obtain, after 
Fourier transformation, 
%%%%%%%%%%%%%%%%%%%%%%%%%%%%%%%%%%%%%%%%%%%%%
\begin{eqnarray} 
\hat{\bf G} (X; P) & \simeq &  \hat{\bf G}_{R A} (X; P) + \hat{M}_+ 
\left[ {\bf G}_R (X; P) - {\bf G}_A (X; P) \right] f (X; P) 
\nonumber \\ 
&& + \hat{M}_+ \left[ {\bf G}_K (X; P) - 
\frac{i}{2} \left\{ {\bf G}_R + {\bf G}_A, \; f 
\right\}_{\mbox{\scriptsize{P.B.}}} \right] 
\, , \nonumber \\ 
\label{3.111} 
\end{eqnarray} 
%%%%%%%%%%%%%%%%%%%%%%%%%%%%%%%%%%%%%%%%%%%%%%
where  
%%%%%%%%%%%%%%%%%%%%%%%%%%%%%%%%%%%%%%%%%%%%%
\begin{equation} 
\hat{\bf G}_{R A} (X; P) \equiv \left( 
\begin{array}{cc} 
{\bf G}_R (X; P) & \;\;\, 0 \\ 
{\bf G}_R (X; P) - {\bf G}_A (X; P) & \;\;\, - {\bf G}_A (X; P) 
\end{array} 
\right) \, . 
\label{312} 
\end{equation} 
%%%%%%%%%%%%%%%%%%%%%%%%%%%%%%%%%%%%%%%%%%%%%%

Here we note that $\hat{\bf \Pi}$ consists of two pieces, 
%%%%%%%%%%%%%%%%%%%%%%%%%%%%%%%%%%%%%%%%%%%%%
\begin{equation} 
\hat{\bf \Pi} = \hat{\bf \Pi}^{\mbox{\scriptsize{loop}}} + \hat{\bf 
\Pi}^{(c)} \, , 
\label{pipi} 
\end{equation} 
%%%%%%%%%%%%%%%%%%%%%%%%%%%%%%%%%%%%%%%%%%%%%
where $\hat{\bf \Pi}^{\mbox{\scriptsize{loop}}}$ is the contribution 
from loop diagrams and $\hat{\bf \Pi}^{(c)}$ is as in 
Eq.~(\ref{sigmac}), which has come from the counter action ${\cal 
A}_c$. It should be remarked that some $\hat{\bf 
\Pi}^{\mbox{\scriptsize{loop}}}$ contains internal vertex(es) $i 
\hat{\bf \Pi}^{(c)}$. [cf. the argument at Sec~VI below.] From the 
above definitions of ${\bf \Pi}_{R (A)}$, ${\bf \Pi}_K$, and 
$\hat{\bf \Pi}^{(c)}$, we have 
%%%%%%%%%%%%%%%%%%%%%%%%%%%%%%%%%%%%%%%%%%%%%
\begin{eqnarray} 
{\bf \Pi}_R^{(c)} & = & {\bf \Pi}_A^{(c)} = 0 \, , \nonumber \\ 
\Pi_K^{(c) \mu \nu} (X; P) & = & - 2 i {\cal P}^{\mu \nu} 
(\hat{\bf p}) P \cdot \partial_X f (X; P) \, . 
\label{56} 
\end{eqnarray} 
%%%%%%%%%%%%%%%%%%%%%%%%%%%%%%%%%%%%%%%%%%%%%

From Eq.~(\ref{3.101}), we obtain, after some manipulation, 
%%%%%%%%%%%%%%%%%%%%%%
\begin{eqnarray} 
{\bf G}_{R (A)} (X; P) & \simeq & {\bf G}_{R (A)}^{(0)} (X; P) + 
{\bf G}_{R (A)}^{(1)} (X; P) \, , \nonumber \\ 
G_{R (A)}^{(0) \mu \nu} (X; P) & = & - \frac{{\cal P}_T^{\mu \nu} 
(\hat{\bf p})}{P^2 - \Pi_{R (A)}^T (X; P)} + \frac{n^\mu n^\nu}{p^2 
+ \Pi_{R (A)}^L (X; P)} \, , 
\label{4.199} 
\\ 
G_{R (A)}^{(1)} (X; P) & = & \frac{i}{2 p^2} \frac{1}{P^2 - \Pi_{R 
(A)}^T} \left[ \frac{1}{P^2 - \Pi_{R (A)}^T} \left( 
p^{\underline{\mu}} \nabla_\perp^{\underline{\nu}} - 
p^{\underline{\nu}} \nabla_\perp^{\underline{\mu}} \right) \Pi_{R 
(A)}^T \right. \nonumber \\ 
&& \mbox{\hspace*{18ex}} \left. - \frac{p_0}{p^2 + \Pi_{R (A)}^L} 
\left( n^\mu \nabla_\perp^{\underline{\nu}} - n^\nu 
\nabla_\perp^{\underline{\mu}} \right) \Pi_{R (A)}^C \right] \, . 
\label{4.21} 
\end{eqnarray} 
%%%%%%%%%%%%%%%%%%%%%%%%%%%%%%%%%%%%
Although $G^{(1)}_{R (A)}$ can be ignored to the gradient 
approximation, we have displayed it. Straightforward 
calculation of the last term in Eq.~(\ref{3.111}) using the 
definitions (\ref{3.101}), (\ref{GK}), and (\ref{PB}) yields (cf. 
Appendix B) 
%%%%%%%%%%%%%%%%%%%%%%%%%%%%%%%%%%%%%%
%%%%%%%%%%%%%%%%%%%%%%%%%%%%%%%%%%%%%%
\[ 
{\bf G}_K (X; P) - \frac{i}{2} \left\{ {\bf G}_R + {\bf G}_A, \, 
\, f \right\}_{\mbox{\scriptsize{P.B.}}} = {\bf G}_K^{(1)} + {\bf 
G}_K^{(2)} \, , 
\] 
%%%%%%%%%%%%%%%%%%%%%%%%%%%%%%%%%%%%%%%%%%%%%%
where 
%%%%%%%%%%%%%%%%%%%%%%%%%%%%%%%%%%%%%%
\begin{eqnarray} 
G_K^{(1) \mu \nu} & = & - i \left[ 2 P \cdot \partial_X f + \left\{ 
\mbox{Re} \Pi_R^T , \;\, f \right\}_{\scriptsize{\mbox{P.B.}}} + i 
(\Pi_K^T)^{\scriptsize{\mbox{loop}}} \right] \frac{{\cal P}_T^{\mu 
\nu}}{(P^2 - \Pi_R^T) (P^2 - \Pi_A^T)} \nonumber \\ 
&& - i \left[ 2 {\bf p} \cdot \nabla_{\bf X} f + \left\{ \mbox{Re} 
\Pi_R^L , \;\, f \right\}_{\scriptsize{\mbox{P.B.}}} + i 
(\Pi_K^L)^{\scriptsize{\mbox{loop}}} \right] \frac{n^\mu n^\nu}{(p^2 
+ \Pi_R^L) (p^2 + \Pi_A^L)} \, , 
\label{3.177} \\ 
G_K^{(2)} & = & - \frac{p_0}{p^2} \mbox{Im} \Pi_R^C \left[ 
\frac{n^\nu \nabla_\perp^{\underline{\mu}} f}{(P^2 - \Pi_R^T) (p^2 + 
\Pi_A^L)} 
- \frac{n^\mu 
\nabla_\perp^{\underline{\nu}} f}{(P^2 - \Pi_A^T) (p^2 + \Pi_R^L)} 
\right] \nonumber \\ 
&& + \frac{i}{2 p} \left( \frac{1}{P^2 - 
\Pi_R^T} - \mbox{c.c.} \right) \left( p^{\underline{\mu}} 
\nabla_\perp^{\underline{\nu}} f - p^{\underline{\nu}} 
\nabla_\perp^{\underline{\mu}} f \right) \nonumber \\ 
&& + i \mbox{Re} \left[ \frac{{\cal P}_T^{\mu 
\nu}}{(P^2 - \Pi_R^T)^2} \left( 2 P \cdot \partial_X f + \left\{ 
\Pi_R^T, \;\, f \right\}_{\scriptsize{\mbox{P.B.}}} \right) 
\right. \nonumber \\ 
&& \mbox{\hspace*{1ex}} \left. \mbox{\hspace*{6ex}} + \frac{n^\mu 
n^\nu}{(p^2 + \Pi_R^L)^2} \left( 2 {\bf p} \cdot \nabla_{\bf X} f + 
\left\{ \Pi_R^L, \;\, f \right\}_{\scriptsize{\mbox{P.B.}}} \right) 
\right] \, , 
\label{GK2} 
\end{eqnarray} 
%%%%%%%%%%%%%%%%%%%%%%%%%%%%%%%%%%%%%%%%%%%%%%%%%%%%%%%%%%%%
where 
%%%%%%%%%%%%%%%%%%%%%%%%%%%%%%%%%%%%%%
\begin{equation} 
\left( \Pi_K^{T (L)} \right)^{\mbox{\scriptsize{loop}}} \equiv [1 + 
f] \left( \Pi_{1 2}^{T (L)} \right)^{\mbox{\scriptsize{loop}}} 
- f \left( \Pi_{2 1}^{T (L)} \right)^{\mbox{\scriptsize{loop}}} 
\, . 
\label{3.188} 
\end{equation} 
%%%%%%%%%%%%%%%%%%%%%%%%%%%%%%%%%%%%%%%%%%%%%%%%%%%%%%%%%%%%
In obtaining Eqs.~(\ref{3.177}) and (\ref{GK2}), use has been made 
of $\left( \Pi_A^{T (L)} \right)^* = \Pi^{T (L)}_R$, which is proved 
in Appendix A. 
%%%%%%%%%%%%%%%%%%%%%%%%%%%%%%%%%%%%%%%%%%%%%%
%%%% SECTION III %%%%%%%%%%%%%%%%%%%%%%%%%%%%%%%
%%%%%%%%%%%%%%%%%%%%%%%%%%%%%%%%%%%%%%%%%%%%%%
\section{Generalized Boltzmann equation} 
\subsection{Energy-shell and physical number densities} 
For later use, referring to Eq.~(\ref{4.199}), we define the 
energy-shell for the transverse mode by 
%%%%%%%%%%%%%%%%%%%%%%
\begin{equation} 
\mbox{Re} \left[ P^2 - \mbox{Re} \Pi_R^T (X; P) \right]_{p^0 = 
\pm \omega_T (X; \pm {\bf p})} = 0 \, . 
\label{D2} 
\end{equation} 
%%%%%%%%%%%%%%%%%%%%%%
It is well known \cite{le-b} that, in equilibrium quark-gluon 
plasma, \lq\lq longitudinal'' mode called plasmon appears for soft 
$p$ $(= O (g T))$ [$g$ the QCD coupling constant]. The energy-shell 
of such modes is defined by [cf. Eq.~(\ref{4.199})] 
%%%%%%%%%%%%%%%%%%%%%%
\begin{equation} 
\mbox{Re} \left[ p^2 + \mbox{Re} \Pi_R^L (X; P) \right]_{p^0 = \pm 
\omega_L (X; \mp {\bf p})} = 0 \, . 
\label{D3} 
\end{equation} 
%%%%%%%%%%%%%%%%%%%%%%
Useful formulae that hold on the energy-shell are displayed in 
Appendix C.  

We restrict to the $p_0 > 0$ part. The $p_0 < 0$ part yields the 
same result, since a gluon and a corresponding antigluon is the 
same. In some of the formulae in this subsection, we shall drop the 
argument $X$. 

For obtaining expressions for the number densities of the transverse 
and \lq\lq longitudinal'' modes, we compute the statistical average 
of the operators, 
%%%%%%%%%%%%%%%%%%%%%%%%%%%%%%%%%%%%%%
\begin{eqnarray*} 
{\cal N}_T (x) & = & - i A^{(+)}_j (x) 
\stackrel{\leftrightarrow}{\frac{\partial}{\partial x_0}} 
A^{(-)}_j (x) \, , \nonumber \\ 
{\cal N}_L (x) & = & i A^{(+)}_0 (x) 
\stackrel{\leftrightarrow}{\frac{\partial}{\partial x_0}} 
A^{(-)}_0 (x) \, . 
\end{eqnarray*} 
%%%%%%%%%%%%%%%%%%%%%%%%%%%%%%%%%%%%%%
Here the superscript \lq\lq $(+)$'' [\lq\lq $(-)$''] stands for the 
positive [negative] frequency part. Statistical averages of ${\cal 
N}_T$ and of ${\cal N}_L$ yield, in respective order, 
%%%%%%%%%%%%%%%%%%%%%%%%%%%%%%%%%%%%%%
\begin{eqnarray} 
\mbox{Tr} \left[ {\cal N}_T (x) \rho \right] & = & i \int 
\frac{d^{\, 4} P}{(2 \pi)^4} \theta (p_0) p_0 \left[ G_{1 
2}^{j j} (X; P) + G_{2 1}^{j j} 
(X; P) \right] \, , 
\nonumber \\ 
\mbox{Tr} \left[ {\cal N}_L (x) \rho \right] & = & - i \int 
\frac{d^{\, 4} P}{(2 \pi)^4} \theta (p_0) p_0 \left[ G_{1 2}^{0 0} 
(X; P) + G_{2 1}^{0 0} (X; P) \right] \, . 
\label{kawa} 
\end{eqnarray} 
%%%%%%%%%%%%%%%%%%%%%%%%%%%%%%%%%%%%%%
It is to be noted that \lq\lq $x$'' here is a macroscopic spacetime 
coordinates [cf. Sec.~I]. 

We first compute the leading contributions to Eq.~(\ref{kawa}). 
Substituting the leading parts of $G_{2 1}$ and of $G_{1 2}$ (cf. 
Eq.~(\ref{3.111})) and using Eq.~(\ref{4.199}), we obtain 
%%%%%%%%%%%%%%%%%%%%%%%%%%%%%%%%%%%%%%%%%%%%%%
\begin{eqnarray*} 
\mbox{Tr} \left[ {\cal N}_T (x) \rho \right] & = & 2 i \int 
\frac{d^{\, 4} P}{(2 \pi)^4} \theta (p_0) p_0 \left[ \frac{1}{P^2 - 
\Pi_R^T} - \mbox{c.c.} \right] [2 f (X; P) + 1] \, , \nonumber \\ 
\mbox{Tr} \left[ {\cal N}_L (x) \rho \right] & = & - i \int 
\frac{d^{\, 4} P}{(2 \pi)^4} \theta (p_0) p_0 \left[ \frac{1}{p^2 + 
\Pi_R^L} - \mbox{c.c.} \right] [2 f (X; P) + 1] \, . 
\end{eqnarray*} 
%%%%%%%%%%%%%%%%%%%%%%%
The narrow-width approximation\footnote{In the case of equilibrium 
system, the narrow-width approximation is a good approximation for 
hard modes but not for soft modes.} $\mbox{Im} \Pi_R^{T (L)} \to 
- \epsilon (p_0) 0^+$ yields 
%%%%%%%%%%%%%%%%%%%%%%%%%%%%%%%%%%
\begin{eqnarray} 
\mbox{Tr} \left[ {\cal N}_T (x) \rho \right] & \simeq & 2 \int 
\frac{d^{\, 3} p}{(2 \pi)^3} Z_T (\omega_T ({\bf p}), {\bf p}) n 
(\omega_T ({\bf p}), {\bf p}) + ... \, , 
\label{5.5d} \\ 
\mbox{Tr} \left[ {\cal N}_L (x) \rho \right] & \simeq & \int 
\frac{d^{\, 3} p}{(2 \pi)^3} Z_L (\omega_L ({\bf p}), {\bf p}) n 
(\omega_L ({\bf p}), {\bf p}) + ... \, . 
\label{dennka} 
\end{eqnarray}
%%%%%%%%%%%%%%%%%%%%%%%
Here $n$ is as in Eq.~(\ref{N-desu}), and \lq $...$' stands for the 
contribution from $2 f + 1 \ni \epsilon (p^0)$ [ cf. 
Eq.~(\ref{f-desu})], which is the vacuum-theory contribution 
corrected by the medium effect. $Z$'s in Eqs.~(\ref{5.5d}) and 
(\ref{dennka}) are the wave-function renormalization factors, 
Eqs.~(\ref{Z1}) and (\ref{C.3d}) in Appendix C. If there are several 
modes, summation should be taken over all modes. The factor \lq $2$' 
on the RHS of Eq.~(\ref{5.5d}) is a reflection of the two 
independent polarizations. 

Thus, we have learned that $n$ is the number density of gluonic 
quasiparticle. Undoing the narrow-width approximation yields 
further corrections to the number densities. 

Let us turn to analyze the contributions from the other parts of 
$\hat{G}$ in Eq.~(\ref{3.111}). Inspection of Eq.~(\ref{kawa}) 
with Eqs.~(\ref{3.111})~-~(\ref{GK2}) shows that 
all but $G_K^{(1)}$, Eq.~(\ref{3.177}), yield well-defined 
corrections to the physical number densities due to the medium 
effect. $G_K^{(1)}$ contains 
%%%%%%%%%%%%%%%%%%%%%%%%%%%%%%%%%%%%
\begin{equation} 
\frac{1}{\left[ P^2 - \Pi_R^T \right] \, \left[ P^2 - 
\left( \Pi_R^T \right)^* \right]} 
\label{A}
\end{equation} 
%%%%%%%%%%%%%%%%%%%%%%%%%%%%%%%%%%%%%
and 
%%%%%%%%%%%%%%%%%%%%%%%%%%%%%%%%%%%%
\begin{equation} 
\frac{1}{\left[ p^2 + \Pi_R^L \right] \left[ p^2 + \left( \Pi_R^L 
\right)^* \right]} \, . 
\label{B}
\end{equation} 
%%%%%%%%%%%%%%%%%%%%%%%%%%%%%%%%%%%%%
In the narrow-width approximation $\mbox{Im} \Pi_R^T \to - \epsilon 
(p_0) 0^+$ [$\mbox{Im} \Pi_R^L \to - \epsilon (p_0) 0^+$], 
Eq.~(\ref{A}) [Eq.~(\ref{B})] develops pinch singularity at $p_0 = 
\omega_T ({\bf p})$ [$p_0 = \omega_L ({\bf p})$] in a complex 
$p^0$-plane. Then the contributions of $G_K^{(1)}$ to 
Eq.~(\ref{kawa}) diverge in this approximation. In practice, 
$\mbox{Im} \Pi_R^{T (L)}$ $(\propto g^2)$ is a small quantity (at 
least for hard modes), so that the contribution, although not 
divergent, is large. This invalidates the perturbative scheme and a 
sort of \lq\lq renormalization'' is necessary for the number 
densities \cite{nie,nie1,nie2}. This observation leads us to 
introduce the condition $G_K^{(1)} = 0$ on the energy-shells: 
%%%%%%%%%%%%%%%%%%%%%%%%%%%%%%%%%%%%%%%%%%%%%%%%%%%%%%%%%%
\begin{eqnarray} 
&& \left[ 2 P \cdot \partial_X f + \left\{ \mbox{Re} \Pi_R^T , \;\, 
f \right\}_{\scriptsize{\mbox{P.B.}}} + i 
(\Pi_K^T)^{\scriptsize{\mbox{loop}}} \right]_{p_0 = \pm \omega_T (X; 
\pm {\bf p})} = 0 \, , 
\label{BTT} \\ 
&& \left[ 2 {\bf p} \cdot \nabla_{\bf X} f + \left\{ \mbox{Re} 
\Pi_R^L , \;\, f \right\}_{\scriptsize{\mbox{P.B.}}} + i 
(\Pi_K^L)^{\scriptsize{\mbox{loop}}} \right]_{p_0 = \pm \omega_L (X; 
\pm {\bf p})} = 0 \, , 
\label{BL} 
\end{eqnarray} 
%%%%%%%%%%%%%%%%%%%%%%%%%%%%%
These serve as determining equations for so far arbitrary $f$. (See 
below, for more details.) Now the above-mentioned large 
contributions, which turn out to the diverging contributions (due to 
the pinch singularities) in the narrow-width approximation, do not 
appear. Thus, the contributions from $G_K^{(1)}$ to Eq.~(\ref{kawa}) 
also yields well-defined corrections to the number densities. 
%%%%%%%%%%%%%%%%%%%%%%%%%%%%%%%%%%%%%%%%%%%%%%%
%%%%%% SUB %%%%%%%%%%%%%%%%%%%%%%%%%%%%%%%%%%%%
%%%%%%%%%%%%%%%%%%%%%%%%%%%%%%%%%%%%%
\subsection{Generalized Boltzmann equation} 
We are now in a position to disclose the physical meaning of 
Eqs.~(\ref{BTT}) and (\ref{BL}). With the help of the formulae in 
Appendix C, Eq.~(\ref{BTT}) with $p_0 \to \omega_T (X; {\bf p})$ 
turns out to 
%%%%%%%%%%%%%%%%%%%%%%%%%%%%%%%%%%%%
\begin{eqnarray} 
&& \left( Z_T (X; P) \right)^{- 1} \left[ \frac{\partial}{\partial 
X^0} + {\bf v}_T (X; {\bf p}) \cdot \frac{\partial}{\partial {\bf 
X}} \right] n (X; P) \nonumber \\ 
&& \mbox{\hspace*{5ex}} + \frac{1}{2 \omega_T (X; {\bf p})} 
\frac{\partial \mbox{Re} \Pi_R^T (X; P)}{\partial X^\mu} 
\frac{\partial n (X; P)}{\partial P_\mu} \nonumber \\ 
&& \mbox{\hspace*{9ex}} = - \frac{i}{2 \omega_T (X; 
{\bf p})} \left( \Pi_K^T \right)^{\mbox{\scriptsize{loop}}} (X; 
P) \, . 
\label{PBB}
\end{eqnarray} 
%%%%%%%%%%%%%%%%%%%%%%%%%%%%%%%%%%%%
Here ${\bf v}_T$ $(\equiv \partial \omega_T (X; {\bf p}) / 
\partial {\bf p})$ is the group velocity of the mode [cf. 
Eq.~(\ref{bui})]. As will be shown in Appendix D, the RHS of 
Eq.~(\ref{PBB}) is related to the net production rate, 
$\Gamma^T_{\mbox{\scriptsize{net p}}}$, 
of the mode $p^0 = \omega_T (X; {\bf p})$. Using Eqs.~(\ref{sei11}) 
and (\ref{ato}), we obtain 
%%%%%%%%%%%%%%%%%%%%%%%%%%%%%%%%%%%%
\begin{eqnarray*} 
&& \left[ \frac{\partial}{\partial X^0} + {\bf v}_T (X; {\bf p}) 
\cdot \frac{\partial}{\partial {\bf X}} \right] n (X; P) 
\nonumber \\ 
&& \mbox{\hspace*{5ex}} 
+ \frac{\partial 
\omega_T (X; {\bf p})}{\partial X^\mu} \frac{\partial n (X; 
P)}{\partial P_\mu} = \Gamma_{\mbox{\scriptsize{net}} \; p}^T (X; 
{\bf p}) \, . 
\end{eqnarray*} 
%%%%%%%%%%%%%%%%%%%%%%%%%%%%%%%%%%%%
This can further be rewritten in the form, 
%%%%%%%%%%%%%%%%%%%%%%%%%%%%%%%%%%%%
\begin{eqnarray} 
&& \left( \frac{d}{d X^0} + {\bf v}_T (X; {\bf p}) \cdot 
\frac{d}{d {\bf X}} \right) n (X; \omega_T (X; {\bf p}), \hat{\bf 
p}) 
\nonumber \\ 
&& \mbox{\hspace*{5ex}} - \frac{\partial \omega_T (X; {\bf 
p})}{\partial {\bf X}} \frac{d n}{d {\bf p}} = 
\Gamma_{\mbox{\scriptsize{net}} \; p}^T (X; {\bf p}) \, . 
\label{Bol1} 
\end{eqnarray} 
%%%%%%%%%%%%%%%%%%%%%%%%%%%%%%%%%%%%
Similarly, Eq.~(\ref{BL}) with $p_0 = \omega_L (X; {\bf p})$ yields 
%%%%%%%%%%%%%%%%%%%%%%%%%%%%%%%%%%%%
\begin{eqnarray} 
&& \left( \frac{d}{d X^0} + {\bf v}_L (X; {\bf p}) \cdot 
\frac{d}{d {\bf X}} \right) n (X; \omega_L (X; {\bf p}), 
{\bf p}) 
\nonumber \\ 
&& \mbox{\hspace*{5ex}} - \frac{\partial \omega_L (X; {\bf 
p})}{\partial {\bf X}} \frac{d n}{d {\bf p}} = 
\Gamma_{\mbox{\scriptsize{net}} \; p}^L (X; {\bf p}) \, . 
\label{Bol2} 
\end{eqnarray} 
%%%%%%%%%%%%%%%%%%%%%%%%%%%%%%%%%%%%
$n$ here is essentially (the main part of) the 
relativistic Wigner function, and Eqs.~(\ref{Bol1}) and (\ref{Bol2}) 
are the generalized relativistic Boltzmann equation for 
gluonic quasiparticles [cf.~\cite{heinz}]. 

For the sake of definiteness, we are taking QCD and have dealt with 
the gluon sector. The quark sector has already been dealt with in 
\cite{nie2}. As far as the gluon sector is concerned to the gradient 
approximation, one can use the \lq\lq free quark'' distribution 
function that subjects to the \lq\lq free Boltzmann equation.'' This 
can be seen as follows: In the gluon sector, the quark distribution 
function enters through the quark propagators involved 
in the gluon self-energy part $\hat{\bf \Pi}$. As observed at the 
beginning of Sec.~IV, $\hat{\bf \Pi}$ is related to the gradient 
part of the gluon propagator. Thus, in calculating $\hat{\bf \Pi}$, 
one can use the $0$th-order expression (of the derivative expansion) 
for the propagators involved in $\hat{\bf \Pi}$. 
%%%%%%%%%%%%%%%%%%%%%%%%%%%%%%%%%%%%%%%%%%%%%%
%%%% SECTION III %%%%%%%%%%%%%%%%%%%%%%%%%%%%%%%
%%%%%%%%%%%%%%%%%%%%%%%%%%%%%%%%%%%%%%%%%%%%%%
\section{Perturbation theory} 
As has been discussed in the preceding section, the propagator in 
the physical-$N$ scheme is free from the pinch singular term (in the 
narrow-width approximation) and then the perturbative calculation of 
some quantity yields \lq\lq healthy'' perturbative corrections. For 
constructing a concrete perturbative scheme, one more step is 
necessary. 

To extend the conditions (\ref{BTT}) and (\ref{BL}) to off the 
energy-shells, we divide $f$ into two pieces [cf. 
Eq.~(\ref{f-desu})] 
%%%%%%%%%%%%%%%%%%%%%%%%%%%%%%%%%%%%
\begin{eqnarray} 
f (X; P) & = & \epsilon (p^0) N (X; P) - \theta (- p^0) \nonumber \\ 
& = & \tilde{f} 
(X; P) + f_0 (X; P) \nonumber \\ 
& = & [\epsilon (p^0) \tilde{N} (X; P) - \theta (- p^0) ] + 
\epsilon (p^0) N_0 (X; P) \, . 
\label{6.1d} 
\end{eqnarray} 
%%%%%%%%%%%%%%%%%%%%%%%%%%%%%%%%%%%%
$f_0$ (and then also $\tilde{f}$) is defined as follows: Let ${\cal 
R}_i (X; {\bf p})$ $(i = 1, 2, ...)$ be a region in a $p^0$-plane 
that includes $i$th energy-shell. We choose ${\cal R}_i (X; {\bf 
p})$, such that, for $i \neq j$, ${\cal R}_i \cap {\cal R}_j = 
\emptyset$.  On each energy-shell, $N_0 (X; P) = N (X; P)$, and, in 
whole $p^0$-region but ${\cal R}_i (X; {\bf p})$ $(i = 1, 2, ...)$, 
$N_0 (X; P)$ vanishes. $\partial N_0 (X; P) / 
\partial X$ and $\partial N_0 (X; P) / \partial P$ exist and $N_0 
(X; P)$ obeys 
%%%%%%%%%%%%%%%%%%%%%%%%%%%%%%%%%%%%%%%%%%%%%%%%%%%%%%%%%%
\begin{eqnarray} 
&& 2 P \cdot \partial_X N_0 + \left\{ \mbox{Re} \Pi^T_R, \; N_0 
\right\}_{\mbox{\scriptsize{P.B.}}} = - i \left( \Pi^T_K 
\right)^{\mbox{\scriptsize{{loop}}}} \, , \nonumber \\ 
&& 2{\bf p} \cdot \nabla_{\bf X} N_0 + \left\{ \mbox{Re} \Pi^L_R, \; 
N_0 \right\}_{\mbox{\scriptsize{P.B.}}} = - i \left( \Pi^L_K 
\right)^{\mbox{\scriptsize{{loop}}}} \, . 
\label{kiso1} 
\end{eqnarray} 
%%%%%%%%%%%%%%%%%%%%%%%%%%%%%
Then, $G_K^{(1) \mu \nu}$ in Eq.~(\ref{3.177}) turns out to 
%%%%%%%%%%%%%%%%%%%%
\begin{eqnarray} 
G_K^{(1) \mu \nu} & = & - i \left[ 2 P \cdot \partial_X \tilde{f} + 
\left\{ \mbox{Re} \Pi_R^T , \;\, \tilde{f} 
\right\}_{\scriptsize{\mbox{P.B.}}} + i 
(\Pi_K^T)^{\scriptsize{\mbox{loop}}} \right] \frac{{\cal P}_T^{\mu 
\nu}}{(P^2 - \Pi_R^T) (P^2 - \Pi_A^T)} \nonumber \\ 
&& - i \left[ 2 {\bf p} \cdot \nabla_{\bf X} 
\tilde{f} + \left\{ \mbox{Re} \Pi_R^L , \;\, \tilde{f} 
\right\}_{\scriptsize{\mbox{P.B.}}} + i 
(\Pi_K^L)^{\scriptsize{\mbox{loop}}} \right] \frac{n^\mu n^\nu}{(p^2 
+ \Pi_R^L) (p^2 + \Pi_A^L)} \, . 
\label{yah} 
\end{eqnarray} 
%%%%%%%%%%%%%%%%%%%%%%%%%%
It is obvious from the above construction that this $G_K^{(1) \mu 
\nu}$ does not possess pinch singularities in narrow-width 
approximation, and thus \lq\lq healthy'' perturbation theory is 
established. 

It is worth mentioning that there is arbitrariness in the choice of 
the regions ${\cal R}_i (X; {\bf p})$ $(i = 1, 2, ...)$. 
Furthermore, the choice of the functional forms of $\tilde{f}$ and 
of $f_0$ is also arbitrary, provided that 
%%%%%%%%%%%%%%%%%%%%%%%%%%%%%%%%%%%
\[ 
\tilde{f} (X^0 = X^0_{in}, {\bf X}; P) + f_0 (X^0= X^0_{in}, {\bf 
X}; P) = f (X^0 = X^0_{in}, {\bf X}; P) \, , 
\] 
%%%%%%%%%%%%%%%%%%%%%%%%%%%%%%% 
where $f (X^0 = X^0_{in}, {\bf X}; P)$ is the initial data with 
$X^0_{in}$ the initial time. As has been discussed at the end of 
Sec.~III, these arbitrariness are not the matter. 

To summarize, to the gradient approximation, the (resummed) 
propagator $\hat{\cal G}^{\mu \nu}$ of the theory is 
%%%%%%%%%%%%%%%%%%%%%%%%%%%%%
\begin{equation} 
\hat{\cal G}^{\mu \nu} = \hat{G}^{\mu \nu} + \hat{\Delta}_1^{\mu 
\nu} + \hat{\Delta}_2^{\mu \nu} + \hat{\Delta}_3^{\mu \nu} \, . 
\label{owao} 
\end{equation} 
%%%%%%%%%%%%%%%%%%%%%%%%%%%%%%
Here $\hat{\Delta}_1^{\mu \nu}$~-~$\hat{\Delta}_3^{\mu \nu}$ are as 
in Eqs.~(\ref{2.666})~-~(\ref{dai1}) and $\hat{G}^{\mu \nu}$ is as 
in Eq.~(\ref{3.111}) with Eqs.~(\ref{312})~-~(\ref{3.188}), provided 
that $G_K^{(1) \mu \nu}$ is given by Eq.~(\ref{yah}). $f$ consists 
of two pieces as in Eq.~(\ref{6.1d}). $f_0$ $(= \epsilon (p^0) N_0)$ 
subjects to Eq.~(\ref{kiso1}), which is to be solved under a given 
initial data. It is to be noted that 
$\hat{\Delta}_1^{\mu \nu}$~-~$\hat{\Delta}_3^{\mu \nu}$ are the 
gradient parts, so that, if one wants, for $f$ in 
$\hat{\Delta}_1^{\mu \nu}$~-~$\hat{\Delta}_3^{\mu \nu}$, one can 
substitute the solution to the \lq\lq free Boltzmann equation,'' 
Eq.~(\ref{ba1}). 

Determination of $f$ or $N$ proceeds order by order in perturbation 
theory to get $f = f_0 + f_1 + f_2 + ...$. Provided that 
perturbative computation is completed up to $(n - 1)$th order, one 
can proceed to $n$th order calculation. $f$ or $N$ enters through 
$\hat{\Delta}^{\mu \nu}_{\scriptsize{\mbox{base}}}$, 
Eq.~(\ref{4.555}), and through $\hat{\Pi}^{(c) \mu \nu}$ $(\propto P 
\cdot \partial f)$, Eq.~(\ref{sigmac}). For $f$ in $\hat{\bf 
\Delta}^{\mu \nu}_{\scriptsize{\mbox{base}}}$ and in $\hat{\bf 
\Pi}^{(c)}$ in Eq.~(\ref{pipi}), we substitute $\sum_{j = 0}^n f_j$, 
while for $f$ in $\hat{\bf \Pi}^{(c)}$ involved in some 
$\hat{\bf \Pi}^{\mbox{\scriptsize{loop}}}$, Eq.~(\ref{pipi}), we use 
appropriate $\sum_{i = 0}^j f_i$ $(j \leq n - 1)$. Then, proceeding 
as above, one can determine $f_n$, and thereby any physical quantity 
may be computed. Here we recall that we are taking the gradient 
approximation. Then, in proceeding to higher orders, one should 
check whether or not the consequences of the calculation of the 
order under consideration is consistent with the gradient 
approximation. One suspects that, in general, only first few 
orders meet this criterion. 

As to the vertex, as has been mentioned at Sec.~II A, the matrix 
vertex $\hat{V}$ is diagonal and its $(2 \, 2)$-component is of 
opposite sign relative to the $(1 \, 1)$-component. It is to be 
noted that the two-point vertex $i L_c^{\mu \nu} (x, y) \hat{M}_-$ 
in Eq.~(\ref{sigmac}) has already been built into $\hat{G}^{\mu \nu}$ 
(cf. Eqs.~(\ref{pipi}) and (\ref{56})) and is absent in the 
perturbative framework using $\hat{\cal G}^{\mu \nu}$ in 
Eq.~(\ref{owao}). 
%%%%%%%%%%%%%%%%%%%%%%%%%%%%%%%%%%%%%%%%%%%%%%
%%%% SECTION III %%%%%%%%%%%%%%%%%%%%%%%%%%%%%%%
%%%%%%%%%%%%%%%%%%%%%%%%%%%%%%%%%%%%%%%%%%%%%%
\section{Summary and outlook} 
In this paper we have dealt with out-of-equilibrium perturbation 
theory for gauge bosons and FP-ghosts in Coulomb gauge. 
The gauge-boson and FP-ghost 
propagators are constructed from first principles. Only 
approximation we have 
employed is the so-called gradient approximation, so that the 
perturbative framework applies to the quasiuniform systems near 
equilibrium or the nonequilibrium quasistationary systems. The 
framework allows us to compute any reaction rates. 

There comes out naturally the generalized Boltzmann equation (GBE) 
that describes the spacetime evolution of the number densities of 
quasiparticles, through which the evolution of the system is 
described. 

The GBE for a quark-gluon plasma (nonequilibrium QCD) is \lq\lq 
directly'' derived in \cite{B-I} [cf., also, \cite{mro}]. The 
\lq\lq derivation'' of the GBE in this paper is quite different 
from that in \cite{B-I}. What we have shown here is that the 
requirement of the absence of large contributions from the 
perturbative framework leads to the GBE. This means that the 
quasiparticles thus defined are the well-defined modes in the 
medium. Conversely, if we start with defining the quasiparticles 
such that their number density functions subject to the GBE, then, 
on the basis of them, well-defined perturbation theory may be 
constructed. 
%%%%%%%%%%%%%%%%%%%%%%%%%%%%
\section*{Acknowledgemente}
This work was supported in part by a Grant-in-Aide for Scientific 
Research ((C)(2) (No.~12640287)) of the Ministry of Education, 
Science, Sports and Culture of Japan. 
%%%%%%%%%%%%%%%%%%%%%%%%%%%%%%%
%%%%%%%%%%%%%%%%%%%%%%%%%%%%%%%
%%%% APP %%%%%%%%%%%%%%%%%%%%%%%%%%%%%%%%%%%%%%%%%%%%%%%%%%%
%%%%%%%%%%%%%%%%%%%%%%%%%%%%%%%%%%%%%%%%%%%%%%%%%%%%%%%%%%%%
\setcounter{equation}{0}
\setcounter{section}{1}
\section*{Appendix A: Properties of the self-energy parts} 
\def\theequation{\mbox{\Alph{section}.\arabic{equation}}}
\subsection*{A.1: Gluon self-energy part} 
Let us start with studying the properties of the propagator. From 
Eqs.~(\ref{prop})~-~(\ref{yatto}) and Eq.~(\ref{add}), we obtain 
%%%%%%%%%%%%%%%%%%%%%%%%%%%%%%%%%%%%%%%%%%%%%%%%%%%%%%%%
\begin{eqnarray*} 
\left[i \Delta_{1 1}^{\mu \nu} (X; P) \right]^* & = & i \Delta_{2 
2}^{\nu \mu} (X; P) = i \Delta_{2 2}^{\mu \nu} (X; - P) \, , 
\nonumber \\ 
\left[i \Delta_{1 2 (2 1)}^{\mu \nu} (X; P) \right]^* & = & i 
\Delta_{1 2 (2 1)}^{\nu \mu} (X; P) = \Delta_{2 1 (1 2)}^{\mu \nu} 
(X; - P) \, , 
\end{eqnarray*} 
%%%%%%%%%%%%%%%%%%%%%%%%%%%%%%%%%%%%%%%%%%%%
where use has been made of $f (X; P) = - \theta (- p_0) + \epsilon 
(p_0) n(X; \epsilon (p_0) {\bf p})$. 

Now, we observe the properties of vertices. As has been mentioned in 
Sec.~IIA and Sec.~VI, the vertex matrix is diagonal and its $(2 \, 
2)$-component is of opposite sign relative to the $(1 \, 
1)$-component, $\hat{V} = \mbox{diag} (v^{(1)}, v^{(2)}) = 
\mbox{diag} (v, - v)$. For QCD, let $\hat{V}_3 = (v_3 (P), - v_3 
(P))$ be a 3-gluon-vertex factor and $\hat{V}_4 = (v_4, - v_4)$ be a 
4-gluon-vertex factor. Noting that $v_3 (P)$ is real and linear in 
$P$, and $v_4$, being independent of $P$, is pure imaginary, we see 
that 
%%%%%%%%%%%%%%%%%%%%%%%%%%%%%%%%%%%%%%%%%%%%%%%%%%%%%%%%
\begin{eqnarray} 
\left( v_3^{(1)} (P) \right)^* & = & v_3 (P) = - v_3 ( - P) = 
v_3^{(2)} (- P) \, , \nonumber \\ 
\left( v_4^{(1)} \right)^* & = & - v_4 = v_4^{(2)} 
\label{bunnk} 
\end{eqnarray} 
%%%%%%%%%%%%%%%%%%%%%%%%%%%%%%%%%%%%%%%%%%%%
holds. There is one more (two-point) vertex (\ref{sigmac}), which 
satisfies 
%%%%%%%%%%%%%%%%%%%%%%%%%%%%%%%
\begin{equation} 
\left[ i L_c^{\mu \nu} (X; P) \hat{M}_- \right]^* = i L_c^{\mu \nu} 
(X; - P) \hat{M}_- \, . 
\label{D.33} 
\end{equation} 
%%%%%%%%%%%%%%%%%%%%%%%%%%%%%%
For quark \cite{nie2} and FP-ghost, one can write down the 
similar relations, 
which we do not display here. 

From the diagrammatic analysis of $\hat{\bf \Pi} (X; P)$ using the 
above relations, one can straightforwardly obtain the relations, 
%%%%%%%%%%%%%%%%%%%%%%%%%%%%%%%%%%%%%%%%%%%%%%%%%%%%%%%%
\begin{eqnarray*} 
\left[\Pi_{1 1}^{\mu \nu} (X; P) \right]^* & = & - \Pi_{2 2}^{\mu 
\nu} (X; - P) = - \Pi_{2 2}^{\nu \mu} (X; P) \, , \\ 
\left[\Pi_{1 2 (2 1)}^{\mu \nu} (X; P) \right]^* & = & - \Pi_{2 1 (1 
2)}^{\mu \nu} (X; - P) = - \Pi_{1 2 (2 1)}^{\nu \mu} (X; P) \, . 
\end{eqnarray*} 
%%%%%%%%%%%%%%%%%%%%%%%%%%%%%%%%%%%%%%%%%%%%
Using the decomposition (\ref{koo}) for the leading parts of 
$\hat{\bf \Pi}$, we obtain 
%%%%%%%%%%%%%%%%%%%%%%%%%%%%%%%%%%%%%%%%%%%%%%%%%%%%%%%%
\begin{eqnarray} 
\left[ \left( \Pi_S (X; P) \right)_{1 1} \right]^* & = & - \left( 
\Pi_S (X; P) \right)_{2 2} 
\, , \nonumber \\ 
\left[ \left( \Pi_S (X; P) \right)_{1 2 (2 1)} \right]^* & = & - 
\left( \Pi_S (X; P) \right)_{1 2 (2 1)} \, , \nonumber \\ 
&& \mbox{\hspace*{9ex}} (S = T, L, C, D) \, . 
\label{saiso} 
\end{eqnarray} 
%%%%%%%%%%%%%%%%%%%%%%%%%%%%%%%%%%%%%%%%%%%%
The last equation shows that $\left( \Pi_S (X; P) \right)_{1 2 (2 
1)}$ are pure imaginary. 

Let us turn to analyze $\hat{\bf \Pi}$ in configuration space: 
%%%%%%%%%%%%%%%%%%%%%%%%%%%%%%%%%%%%%%%%%%%%%%%%%%%%%%%%%%%%%%%
\begin{eqnarray} 
\hat{\bf \Pi} (x, y) & = & \int \prod_{i = 1, \, 2, ...} \left[ 
d z_i \, d u_i \, d v_i \right] \nonumber \\ 
&& \mbox{\hspace*{5ex}} \times \hat{\bf \Pi} (x, y; z_1, z_2, ... ; 
(u_1, v_1), (u_2, v_2), 
...) \, , 
\label{kai} 
\end{eqnarray} 
%%%%%%%%%%%%%%%%%%%%%%%%%%%%%%%%%%%%%%%%%%%%%%%%%%%%%%%%%
where $x$ and $y$ are the spacetime points to which the external 
legs are attached, $z_i$ $(i = 1, 2, ...)$ are the internal 
vertex-points, and $(u_i, v_i)$ [$i = 1, 2, ...$] are the sets of 
internal vertex-points of $i L_c (u_i, v_i) \hat{M}_-$. From 
Eq.~(\ref{kore}), which leads to Eqs.~(\ref{iti})~-~(\ref{yatto}), 
and Eq.~(\ref{add}), we see that 
%%%%%%%%%%%%%%%%%%%%%%%%%%%%%%%%%%%%%%%%
\begin{equation} 
{\bf \Delta}_{1 i} (x, y) = {\bf \Delta}_{2 i} (x, y) \;\;\;\; 
\mbox{for} \; x^0 > y^0 \;\;\;\;\;\; (i = 1, 2) \, . 
\label{large} 
\end{equation} 
%%%%%%%%%%%%%%%%%%%%%%%%%%%%%%%%%%%%%%%%%
Here we use the \lq\lq largest-time technique'' \cite{kobes}. 
Using the 
relation (\ref{large}) and Eq.~(\ref{bunnk}) (in configuration 
space) together with their quark and FP-ghost counterparts, one can 
show that, in Eq.~(\ref{kai}), when $z_i^0$ of $(z_1^0, z_2^0, ...)$ 
is the largest time, then $\hat{\bf \Pi} (x, y) = 0$. As can be 
shown by using Eq.~(\ref{D.33}) in configuration space, this is also 
the case for the largest $u_i^0$ or for the largest $v_i^0$ of 
[$(u_1^0, v_1^0), (u_2^0, v_2^0), ...$]. Then, in Eq.~(\ref{kai}), 
the largest time is $x^0$ or $y^0$. 

Similar argument as above shows that 
%%%%%%%%%%%%%%%%%%%%%%%%%%%%%%%%%%%%%%%%%%%%%%%%%%%%%%%%
\begin{eqnarray} 
{\bf \Pi}_{2 1} (x, y) & = & - {\bf \Pi}_{1 1} (x, y) \, , \;\; 
{\bf \Pi}_{2 2} (x, y) = - {\bf \Pi}_{1 2} (x, y) 
\mbox{\hspace*{7ex}} \mbox{for} \; x^0 > y^0 \, , 
\nonumber \\ 
{\bf \Pi}_{1 1} (x, y) & = & - {\bf \Pi}_{1 2} (x, y) \, , \;\; 
{\bf \Pi}_{2 1} (x, y) = - {\bf \Pi}_{2 2} (x, y) \, , 
\mbox{\hspace*{7ex}} \mbox{for} \; x^0 < y^0 \, . 
\label{Foo} 
\end{eqnarray} 
%%%%%%%%%%%%%%%%%%%%%%%%%%%%%%%%%%%%%%%%%%%%
From the first (second) relation, we get ${\bf \Pi}_A (x, y)$ $[ = 
{\bf \Pi}_{1 1} + {\bf \Pi}_{2 1}] = 0$ for $x^0 > y^0$, and 
${\bf \Pi}_R (x, y)$ $[ = {\bf \Pi}_{1 1} + {\bf \Pi}_{1 2}] = 0$ 
for $x^0 < y^0$, as they should be. From Eq.~(\ref{Foo}), follows 
%%%%%%%%%%%%%%%%%%%%%%%%%%%%%%%%%%%%%%%%%%%%%%%%%%%%%%%%%%%%%%%
\begin{equation} 
\sum_{i, \, j = 1}^2 {\bf \Pi}_{i j} (x, y) = 0 \, . 
\label{jiko0} 
\end{equation} 
%%%%%%%%%%%%%%%%%%%%%%%%%%%%%%%%%%%%%%%%%%%%%%%%%%%%%%%%%
Using the decomposition (\ref{koo}), after Fourier transformation, 
we obtain 
%%%%%%%%%%%%%%%%%%%%%%%%%%%%%%%%%%%%%%%%%%%%%%%%%%%%%%%%%%%%%%%
\begin{equation} 
\sum_{i, \, j = 1}^2 \left( \Pi_S (X; P) \right)_{i j} = 0 
\;\;\;\;\;\;\; (S = T, L, C, D) \, . 
\label{saso} 
\end{equation} 
%%%%%%%%%%%%%%%%%%%%%%%%%%%%%%%%%%%%%%%%%%%%%%%%%%%%%%%%%
The relation (\ref{saiso}) leads to 
%%%%%%%%%%%%%%%%%%%%%%%%%%%%%%%%%%%%%%%%%%%%%%%%%%%%%%%%%%%%%%%
\begin{equation} 
\left[ \Pi_A^S (X; P) \right]^* = \Pi_R^S (X; P) \;\;\;\;\;\;\; (S = 
T, L, C, D) \, . 
\label{ccc} 
\end{equation} 
%%%%%%%%%%%%%%%%%%%%%%%%%%%%%%%%%%%%%%%%%%%%%%%%%%%%%%%%%
\subsection*{A.2: FP-ghost self-energy part} 
In every diagram representing the FP-ghost self-energy part 
$\hat{\Pi}_g$, two external FP-ghost lines are connected with each 
other. Then, the fact that the bare FP-ghost propagator is diagonal 
[cf. Eq.~(\ref{kore12})] tells us that the matrix self-energy part 
$\hat{\Pi}_g$ is also diagonal, $\hat{\Pi}_g = \mbox{diag} 
(\Pi_g^{(1)}, \Pi_g^{(2)})$. 

In a similar manner to the above subsection, we obtain [cf. 
Eqs~(\ref{saso}) and (\ref{ccc})] 
%%%%%%%%%%%%%%%%%%%%%%%%%%%%%%%%%%%%%%%%%%%%%%%%%%%%%%%%
\begin{eqnarray*} 
&& \Pi_g^{(1)} (X; P) + \Pi_g^{(2)} (X; P) = 0 \, , \nonumber \\ 
&& \left[ \Pi_g^{(1)} (X; P) \right]^* = - \Pi_g^{(2)} (X; P) \, . 
\end{eqnarray*} 
%%%%%%%%%%%%%%%%%%%%%%%%%%%%%%%%%%%%%%%%%%%%
Then, we see that 
%%%%%%%%%%%%%%%%%%%%%%%%%%%%%%%%%%%%%%%%%%%%%%%%%%%%%%%%%%%%%%%
\[ 
\hat{\Pi}_g (X; P) = \hat{\tau}_3 \Pi_g (X; P) \, , \:\:\:\:\: 
\mbox{Im} \Pi_g (X; P) = 0 \, . 
\] 
%%%%%%%%%%%%%%%%%%%%%%%%%%%%%%%%%%%%%%%%%%%%%%%%%%%%%%%%%
%%%%%%%%%%%%%%%%%%%%%%%%%%%%%%%%%%%%%%%%%%%%%%%%%%%%%%%%%%%%
%%%% APP %%%%%%%%%%%%%%%%%%%%%%%%%%%%%%%%%%%%%%%%%%%%%%%%%%%
%%%%%%%%%%%%%%%%%%%%%%%%%%%%%%%%%%%%%%%%%%%%%%%%%%%%%%%%%%%%
\setcounter{equation}{0}
\setcounter{section}{2}
\section*{Appendix B: On the derivation of ${\bf G}_K$ in 
Eqs.~(\ref{3.177}) and (\ref{GK2})} 
Here we deal with a part of ${\bf G}_K$ in Eq.~(\ref{GK}), 
%%%%%%%%%%%%%%%%%%%%%%%%%%%%%%%%%%%%%%%%%%%%%%%%%%%%%%%%%%%%%%%%%
\begin{eqnarray} 
&& \left[ {\bf G}_R \cdot {\bf \Delta}_R^{- 1} \cdot {\bf \Delta}_K 
\cdot {\bf \Delta}_A^{- 1} \cdot {\bf G}_A \right] (x, y) \simeq 
\int \frac{d^{\, 4} P}{(2 \pi)^4} e^{- P \cdot (x - y)} G_K^{'} (X; 
P) \, , \nonumber \\ 
&& G_K^{'} (X; P) = {\bf G}_R (X; P) {\bf \Delta}_R^{- 1} (P) {\bf 
\Delta}_K (X; P) {\bf \Delta}_A^{- 1} (P) {\bf G}_A (X; P) \, , 
\label{yama} 
\end{eqnarray} 
%%%%%%%%%%%%%%%%%%%%%%%%%%%
which is valid within the approximation under consideration. We are 
adopting the strict Coulomb gauge $\alpha = 0$. Since $(\Delta_R^{- 
1} (P))^{\mu \nu} = (\Delta_A^{- 1} (P))^{\mu \nu} = - {\cal D}^{\mu 
\nu} (P)$ [cf. Eq.~(\ref{4.5d})] contains a term $- g^{\mu i} g^{\nu 
j} p^i p^j / \alpha$ $(\alpha \to 0)$, care should be taken for 
computing Eq.~(\ref{yama}). We need $O (\alpha)$ contribution to 
${\bf G}_{R (A)}$. For the present purpose, it is sufficient to 
obtain the leading-order expression, which is obtained from 
Eq.~(\ref{3.101}): 
%%%%%%%%%%%%%%%%%%%%%%%%%%%%%%%%%%%%%%%%%%%%%%%%%%%%%%%%
\begin{eqnarray*} 
G^{\mu \nu}_{R (A)} (X; P) & \simeq & G^{(0) \mu \nu}_{R (A)} (X; P) 
+ \alpha G^{' \, \mu \nu}_{R (A)} (X; P) \, , \nonumber \\ 
G^{' \, \mu \nu}_{R (A)} (X; P) & = & \frac{1}{p^2} 
\left[ 
\frac{p_0^2}{p^2} 
\left( 
\frac{p^2 + \Pi_{R (A)}^C}{p^2 + \Pi_{R (A)}^L} \right)^2 + 
\frac{p_0}{p} (\hat{p}^{\underline{\mu}} n^\nu + 
n^\mu \hat{p}^{\underline{\nu}}) 
\left( 
\frac{p^2 + \Pi_{R (A)}^C}{p^2 + \Pi_{R (A)}^L} \right) + 
\hat{p}^{\underline{\mu}} \hat{p}^{\underline{\nu}} \right] \, , 
\end{eqnarray*} 
%%%%%%%%%%%%%%%%%%%%%%%%%%%%%%%%%%%%%%%%%%%%
where $G^{(0) \mu \nu}_{R (A)}$ is as in Eq.~(\ref{4.199}) and 
$\hat{p}^{\underline{\mu}} \equiv (0, \hat{\bf p})$ as in 
Eq.~(\ref{koo}). 

We observe that 
%%%%%%%%%%%%%%%%%%%%%%%%%%%%%%%%%%%%%%%%%%%%%%%%%%%%%%%%
\begin{eqnarray*} 
G_R^{(0) \, \mu \nu} p_{\underline{\nu}} & = & p_{\underline{\nu}} 
G_A^{(0) \, \nu \mu} = 0 \, , \\ 
p_{\underline{\rho}} G^{' \, \rho \nu}_A & = & G^{' \, \nu \rho}_R 
p_{\underline{\rho}} = \frac{1}{p^2} \left[ p^{\underline{\nu}} + 
\frac{p^2 + \Pi_{R (A)}^C}{p^2 + \Pi_{R (A)}^L} p_0 n^\nu \right] 
\, . 
\end{eqnarray*} 
%%%%%%%%%%%%%%%%%%%%%%%%%%%%%%%%%%%%%%%%%%%%
Substituting these and Eqs.~(\ref{4.199}) and (\ref{itt}) into 
Eq.~(\ref{yama}), we obtain, after some manipulations, 
%%%%%%%%%%%%%%%%%%%%%%%%%%%%%%%%%%%%%%%%%%%%%%%%%%%%%%%%
\begin{eqnarray*} 
G_K^{'} (X; P) & = & - \frac{i}{p^2} \left[ \frac{2 p^2 n^\mu 
n^\nu}{(p^2 + \Pi_R^L) (p^2 + \Pi_A^L) } {\bf p} \cdot \nabla_X f + 
\frac{p^{\underline{\nu}} \nabla_\perp^{\underline{\mu}} f}{P^2 - 
\Pi_R^T} \right. \nonumber \\ 
&& \left. + \frac{p^{\underline{\mu}} \nabla_\perp^{\underline{\nu}} 
f}{P^2 - \Pi_A^T} + p_0 \left( \frac{\Pi_R^C n^\mu 
\nabla_\perp^{\underline{\nu}}}{(p^2 + \Pi_R^L)(P^2 - \Pi_A^T)} + 
\frac{\Pi_A^C n^\nu \nabla_\perp^{\underline{\mu}}}{(p^2 + 
\Pi_A^L)(P^2 - \Pi_R^T)} \right) \right] \, . 
\end{eqnarray*} 
%%%%%%%%%%%%%%%%%%%%%%%%%%%%%%%%%%%%%%%%%%%%
%%%%%%%%%%%%%%%%%%%%%%%%%%%%%%%%%%%%%%%%%%%%%%%%%%%%%%%%%
%%%% APP %%%%%%%%%%%%%%%%%%%%%%%%%%%%%%%%%%%%%%%%%%%%%%%%%%%
%%%%%%%%%%%%%%%%%%%%%%%%%%%%%%%%%%%%%%%%%%%%%%%%%%%%%%%%%%%%
\setcounter{equation}{0}
\setcounter{section}{3}
\section*{Appendix C: On the energy shell
} 
\def\theequation{\mbox{\Alph{section}.\arabic{equation}}}
Here we display some formulae, which hold on the energy-shells of 
quasiparticles [cf. Eqs.~(\ref{D2}) and (\ref{D3}) with 
Eq.~(\ref{4.199})]. In most formulae in this 
Appendix, the argument $X$ is dropped. We restrict to the positive 
$p_0$ part. 

\noindent {\em Transverse modes}: 

We first define the wave-function renormalization factors through 
taking derivative of Eq.~(\ref{D2}) with respect to $p^0$: 
%%%%%%%%%%%%%%%%%%%%%%
\begin{equation} 
\left( Z_T (\omega_T ({\bf p}), {\bf p}) \right)^{- 1} = 1 - 
\frac{1}{2 \omega_T} \frac{\partial \, \mbox{Re} \, \Pi_R^T (p^0, 
{\bf p})}{\partial p^0} \, \rule[-3mm]{.14mm}{8.5mm} 
\raisebox{-2.85mm}{\scriptsize{$\; p^0 = \omega_T ({\bf 
p})$}} \, . 
\label{Z1} 
\end{equation} 
%%%%%%%%%%%%%%%%%%%%%%
The group velocities of the modes are obtained from the definition 
(\ref{D2}): 
%%%%%%%%%%%%%%%%%%%%%%
\begin{eqnarray} 
{\bf v}_T ({\bf p}) & \equiv & \frac{d \omega_T ({\bf p})}{d {\bf 
p}} = \frac{Z_T (\omega_T ({\bf p}), {\bf p})}{\omega_T ({\bf p})} 
\nonumber \\ 
&& \times \left[ {\bf p} + \frac{1}{2} 
\frac{\partial \, \mbox{Re} \, \Pi_R^T (p^0, {\bf p})}{\partial 
{\bf p}} \, \rule[-3mm]{.14mm}{8.5mm} 
\raisebox{-2.85mm}{\scriptsize{$\; p^0 = \omega_T ({\bf 
p})$}} \right] \, . \nonumber \\ 
\label{bui} 
\end{eqnarray} 
%%%%%%%%%%%%%%%%%%%%%%
Differentiation of Eq.~(\ref{D2}) with respect to $X$ leads to 
%%%%%%%%%%%%%%%%%%%%%%
\begin{equation} 
\frac{\partial \omega_T (X; {\bf p})}{\partial X} = \frac{Z_T 
(\omega_T ({\bf p}), {\bf p})}{2 \omega_T} \frac{\partial \, 
\mbox{Re} \, \Pi_R^T (X; \omega_T (X; {\bf p}), {\bf p})}{\partial 
X} \, . 
\label{ato} 
\end{equation} 
%%%%%%%%%%%%%%%%%%%%%%

\noindent {\em \lq\lq Longitudinal'' modes}: 

From Eq.~(\ref{D3}), we obtain 
%%%%%%%%%%%%%%%%%%%%%%
\begin{eqnarray} 
&& \left( Z_L (\omega_L ({\bf p}), {\bf p}) \right)^{- 1} 
= \frac{1}{2 \omega_L} \frac{\partial \, \mbox{Re} \, \Pi_R^L 
(p^0, {\bf p})}{\partial p^0} \, \rule[-3mm]{.14mm}{8.5mm} 
\raisebox{-2.85mm}{\scriptsize{$\; p^0 = \omega_L ({\bf 
p})$}} \, , \nonumber \\ 
\label{C.3d} \\ 
&& {\bf v}_L ({\bf p}) \equiv \frac{d \omega_L ({\bf p})}{d {\bf p}} 
\nonumber \\  
&& \mbox{\hspace*{6.3ex}} = - \frac{Z_L (\omega_L ({\bf p}), 
{\bf p})}{\omega_L ({\bf p})} \left[ {\bf p} + 
\frac{1}{2} \frac{\partial \, \mbox{Re} \, \Pi_R^L (p^0, 
{\bf p})}{\partial {\bf p}} \, 
\rule[-3mm]{.14mm}{8.5mm} \raisebox{-2.85mm}{\scriptsize{$\; p^0 = 
\omega_L ({\bf p})$}} \right] \, . \nonumber 
\end{eqnarray} 
%%%%%%%%%%%%%%%%%%%%%%
%%%%%%%%%%%%%%%%%%%%%%%%%%%%%%%%%%%%%%%%%%%%%%%%%%%%%%%%%%%%
%%%% APP %%%%%%%%%%%%%%%%%%%%%%%%%%%%%%%%%%%%%%%%%%%%%%%%%%%
%%%%%%%%%%%%%%%%%%%%%%%%%%%%%%%%%%%%%%%%%%%%%%%%%%%
\setcounter{equation}{0}
\setcounter{section}{4}
\section*{Appendix D: Net production rates} 
From Eqs.~(\ref{4.199}), (\ref{D2}) and (\ref{D3}), we see that the 
projection operators onto $p^0 = \omega_T$ $(p^0 = \omega_L)$ mode 
is ${\cal P}_T^{\mu \nu} (\hat{\bf p})$ $(n^\mu n^\nu)$. Then, the 
production (decay) rates of the transverse modes and of the \lq\lq 
longitudinal'' modes are written as \cite{chou,nie,nie1,nie2} 
%%%%%%%%%%%%%%%%%%%%%%%%%%%%%%%%%%%%
\begin{eqnarray*} 
\Gamma_{p (d)}^T ({\bf p}) & = & \frac{i}{2 \omega_T ({\bf p})} 
Z_T (\omega_T ({\bf p}), {\bf p}) \frac{1}{2} \Pi_{1 2 (2 1)}^{\mu 
\nu} ({\cal P}_T (\hat{\bf p}))_{\mu \nu} \nonumber \\ 
& = & \frac{i}{2 \omega_T} Z_T \Pi_{1 2 (2 1)}^T 
\end{eqnarray*} 
%%%%%%%%%%%%%%%%%%%%%%%%%%%%%%%%%%%%
and 
%%%%%%%%%%%%%%%%%%%%%%%%%%%%%%%%%%%%
\begin{eqnarray*} 
\Gamma_{p (d)}^L ({\bf p}) & = & \frac{i}{2 \omega_L ({\bf p})} 
Z_L (\omega_L ({\bf p}), {\bf p}) \Pi_{1 2 (2 1)}^{\mu \nu} n_\mu 
n_\nu \nonumber \\ 
& = & \frac{i}{2 \omega_L} Z_L \Pi_{1 2 (2 1)}^L \, , 
\end{eqnarray*} 
%%%%%%%%%%%%%%%%%%%%%%%%%%%%%%%%%%%%
respectively. Here $Z$'s are the wave-function renormalization 
factor, Eqs.~(\ref{Z1}) and (\ref{C.3d}). 
Thus, the net production rate is 
%%%%%%%%%%%%%%%%%%%%%%%%%%%%%%%%%%%%
\begin{eqnarray} 
\Gamma_{\mbox{\scriptsize{net}} \; p}^{T (L)} ({\bf p}) & = & [1 + 
n (\omega_{T (L)} ({\bf p}), \hat{\bf p})] \Gamma_p^{T (L)} (X; {\bf 
p}) - n (\omega_{T (L)} ({\bf p}), \hat{\bf p}) \Gamma_d^{T (L)} 
({\bf p}) \nonumber \\ 
& = & \frac{i}{2 \omega_{T (L)}} Z_{T (L)} (\omega_{T (L}) 
({\bf p}), {\bf p}) \left( \Pi_K^{T (L)} (\omega_{T (L)} (X; {\bf 
p}), {\bf p}) \right)^{\mbox{\scriptsize{loop}}} \, , 
\label{sei11} 
\end{eqnarray} 
%%%%%%%%%%%%%%%%%%%%%%%%%%%%%%%%%%%%
where $\left( \Pi^{T (L)}_K \right)^{\mbox{\scriptsize{loop}}}$ is 
as in Eq.~(\ref{3.188}). 
%%%%%%%%%%%%%%%%%%%%%%%%%%%%%%%%%%
%%%%%% REFERENCES %%%%%%%%%%%%%%%%
%%%%%%%%%%%%%%%%%%%%%%%%%%%%%%%%%%
 

\newpage 
\begin{thebibliography}{99}
%%%%%%%%%%%%%%%%%%%%%%%%%%%%%%%%%%%%%%%%%%%%%%
\bibitem{bjo} J. D. Bjorken, {\it Acta Phys. Polon. B} {\bf 28}, 
2773 (1997); See also, e.g., K. Rajagopal, in Quark-Gluon Plasma 2, 
edited by R. C. Hwa (World Scientific, Singapore, 1995), p.484. 
%%%%%%%%%%%%%%%%%%%%%%%%%%%%%%%%%%%%%%%%%%
\bibitem{le-b} M. Le Bellac, \lq\lq Thermal Field Theory,'' 
Cambridge University Press, Cambridge, England, 1996. 
%%%%%%%%%%%%%%%%%%%%%%%%%%%%%%%%%%
\bibitem{nie} A.~Ni\'egawa, Prog. Theor. Phys. {\bf 102}, 1 (1999). 
%%%%%%%%%%%%%%%%%%%%%%%%%%%%%%%%%%
\bibitem{nie1} A.~Ni\'egawa, Phys. Rev. D {\bf 62}, 125004 (2000). 
%%%%%%%%%%%%%%%%%%%%%%%%%%%%%%%%%%
\bibitem{nie2} A.~Ni\'egawa, hep-ph/0101216. 
%%%%%%%%%%%%%%%%
\bibitem{sch} 
J. Schwinger, J. Math. Phys. {\bf 2}, 407 (1961); L. V. Keldysh, Zh. 
Eksp. Teor. Fiz. {\bf 47}, 1515 (1964) (Sov. Phys. JETP {\bf 20}, 
1018 (1965)); R. A. Craig, J. Math. Phys. {\bf 9}, 605 (1968); J. 
Rammer and H. Smith, Rev. Mod. Phys. {\bf 58}, 323 (1986). 
%%%%%%%%%%%
\bibitem{chou} K.-C. Chou, Z.-B. Su, B.-L. Hao, and L. Yu, 
Phys. Rep. {\bf 118}, 1 (1985). 
%%%%%%%%%%%
\bibitem{lan} N. P. Landsman and Ch. G. van Weert, Phys. Rep. 
{\bf 145}, 141 (1987). 
%%%%%%%%%%%
\bibitem{noo} A. Ni\'egawa, K. Okano, and H. Ozaki, Phys. Rev. D 
{\bf 61}, 056004 (2000). For the case of equilibrium systems, see, 
A. Ni\'egawa, {\it ibid}. {\bf 57}, 1379 (1998). 
%%%%%%%%%%%%%%%%%
\bibitem{ume} H.~Umezawa, \lq\lq Advanced Field Theory --- Micro, 
Macro, and Thermal Physics.'' AIP, New York, 1993; Y. Yamanaka, H. 
Umezawa, K. Nakamura, and T. Arimitsu, Int. J. Mod. Phys. {\bf A9}, 
1153 (1994); Y. Yamanaka and K. Nakamura, Mod. Phys. Lett. {\bf A9}, 
2879 (1994); H. Chu and H. Umezawa, Int. J. Mod. Phys. {\bf A9}, 
1703 and 2363 (1994). 
%%%%%%%%%%%%%%%%%%%%%
\bibitem{heinz} P. Zhuang and U. Heinz, Phys. Rev. D {\bf 57}, 6525 
(1998); S. Ochs and U. Heinz, Ann. Phys. (N.Y.) {\bf 266}, 351 
(1998). 
%%%%%%%%%%%%%%%%%%%%%%%
\bibitem{B-I} J.-P.~Blaizot and E.~Iancu, Nucl. Phys. {\bf B557}, 
183 (1999). 
\bibitem{mro} S.~ Mr\'owczy\'nski, Phys. Part. Nucl. {\bf 30}, 419 
(1999) [hep-ph/9805435], and earlier works quoted therein. 
\bibitem{kobes} R.~Kobes, Phys. Rev. D {\bf 42}, 562 (1990). 
\end{thebibliography}
\end{document}